\documentclass[aps,prl,reprint,noeprint,showpacs,showkeys,superscriptaddress,longbibliography,floatfix]{revtex4-2}
\usepackage[utf8]{inputenc}
\usepackage{cmap} 
\usepackage[T1]{fontenc}
\usepackage{amsmath,amssymb,mathtools}
\usepackage{hyperref}
\usepackage[dvipsnames]{xcolor}
\hypersetup{
    colorlinks,
    linkcolor={blue!80!black},
    citecolor={blue!80!black},
    urlcolor={blue!80!black}
}
\usepackage{graphicx}
\usepackage{tikz}
\usetikzlibrary{calc}
\usetikzlibrary{math}
\begin{document}

\title{Superdiffusion-like behavior in zero-temperature coarsening of the $d=3$ Ising model}
\author{Denis Gessert}
\email{denis.gessert@itp.uni-leipzig.de}
\affiliation{Institut f\"ur Theoretische Physik, Universit\"at Leipzig, IPF 231101, 04081 Leipzig, Germany}
\affiliation{Centre for Fluid and Complex Systems, Coventry University, Coventry CV1~5FB, United Kingdom}
\author{Henrik Christiansen}
\email{henrik.christiansen@itp.uni-leipzig.de}
\altaffiliation{Present address: NEC Laboratories Europe GmbH, Kurfürsten-Anlage 36, 69115 Heidelberg, Germany.}
\affiliation{Institut f\"ur Theoretische Physik, Universit\"at Leipzig, IPF 231101, 04081 Leipzig, Germany}
\author{Wolfhard Janke}
\email{wolfhard.janke@itp.uni-leipzig.de}
\affiliation{Institut f\"ur Theoretische Physik, Universit\"at Leipzig, IPF 231101, 04081 Leipzig, Germany}
\date{\today}

\begin{abstract}
  One key aspect of coarsening following a quench below the critical temperature is domain growth. 
  For the non-conserved Ising model a power-law growth of domains of like spins with exponent $\alpha = 1/2$ is predicted.
  Including recent work, it was not possible to clearly observe this growth law in the special case of a zero-temperature quench in the three-dimensional model.
  Instead a slower growth with $\alpha<1/2$ was reported.
  We attempt to clarify this discrepancy by running large-scale Monte Carlo simulations of lattice sizes up to $L=2048$ employing an efficient GPU implementation.
  Indeed, at late times we measure domain sizes compatible with the expected growth law --
  but surprisingly, at still later times domains even grow superdiffusively, i.e., with $\alpha > 1/2$.
  We argue that this new problem is possibly caused by sponge-like structures emerging at early times.
\end{abstract}
\maketitle

\section{Introduction}
To quantify the kinetics of coarsening processes, i.e., their time evolution from a disordered to the preferred equilibrium state at low temperature, is of major interest in many physical systems~\cite{bray2002theory,Puri_book}.
The studied systems have become more and more complex over the last decades ranging from investigations of interface growth~\cite{ohta1982universal,durang2017} over systems with long-range interactions~\cite{christiansen2018,corberi2019one} to the application of the methods to the study of the collapse dynamics of polymers~\cite{majumder2017SM,christiansen2017JCP}.
Of technical relevance is this process for example in the fabrication of glasses~\cite{bever1985encyclopedia}.

The theory based on deterministic continuum models, predicts for $d$-dimensional systems with short-range interactions and non-conserved $O(n)$ models for all quench temperatures $T$ below the critical temperature $T_c$ a power-law growth of the characteristic length scale of the coarsening domain patterns,
\begin{equation}
  \ell(t) \sim t^{\alpha}
\end{equation}
with $\alpha=1/2$ for all systems with ${d>n}$ or ${n>2}$~\cite{bray2002theory}.
This was confirmed in numerous simulation studies of quenches to $T\neq 0<T_c$ for such models~\cite{bray2002theory,Puri_book,livi2017nonequilibrium}.
Also in experiments of the ordering kinetics in Cu$_3$Au~\cite{shannon1992time} and at the isotropic-to-cholesteric liquid crystal transition~\cite{dierking2000domain}, a value close to $\alpha=1/2$ was reported. For the special case of a quench to $T=0$ in the $d=2$ Ising model a power law with growth exponent $\alpha=1/2$ is observed as well~\cite{shore1992logarithmically,lipowski1999anomalous}.
\par
Somewhat as a surprise, for a long time, numerical simulations of the coarsening in the $d=3$ Ising model when quenched to zero temperature only reported anomalously small values of $\alpha < 1/2$~\cite{newmanbarkema}, even though many numerical studies were conducted studying the properties of this system~\cite{shore1992logarithmically,cueille1997spin,lipowski1999anomalous,spirin2001fate,spirin2001freezing,olejarz2011zerofreeze,olejarz2011zero}.
Often reported are values of $\alpha \approx 1/3$~\cite{shore1992logarithmically,cueille1997spin,lipowski1999anomalous} when using system sizes of up to $L=240$.
This lower exponent has been attempted to be explained in various ways.
One such attempt targeted on finding arguments and physical explanations for this phenomenon through the fact that the initial $T=\infty$ structure does percolate in three dimensions but not in two dimensions~\cite{lipowski1999anomalous}.
\par

\begin{figure}
  \begin{tabular}{ccc}
    \includegraphics[width=0.15\textwidth]{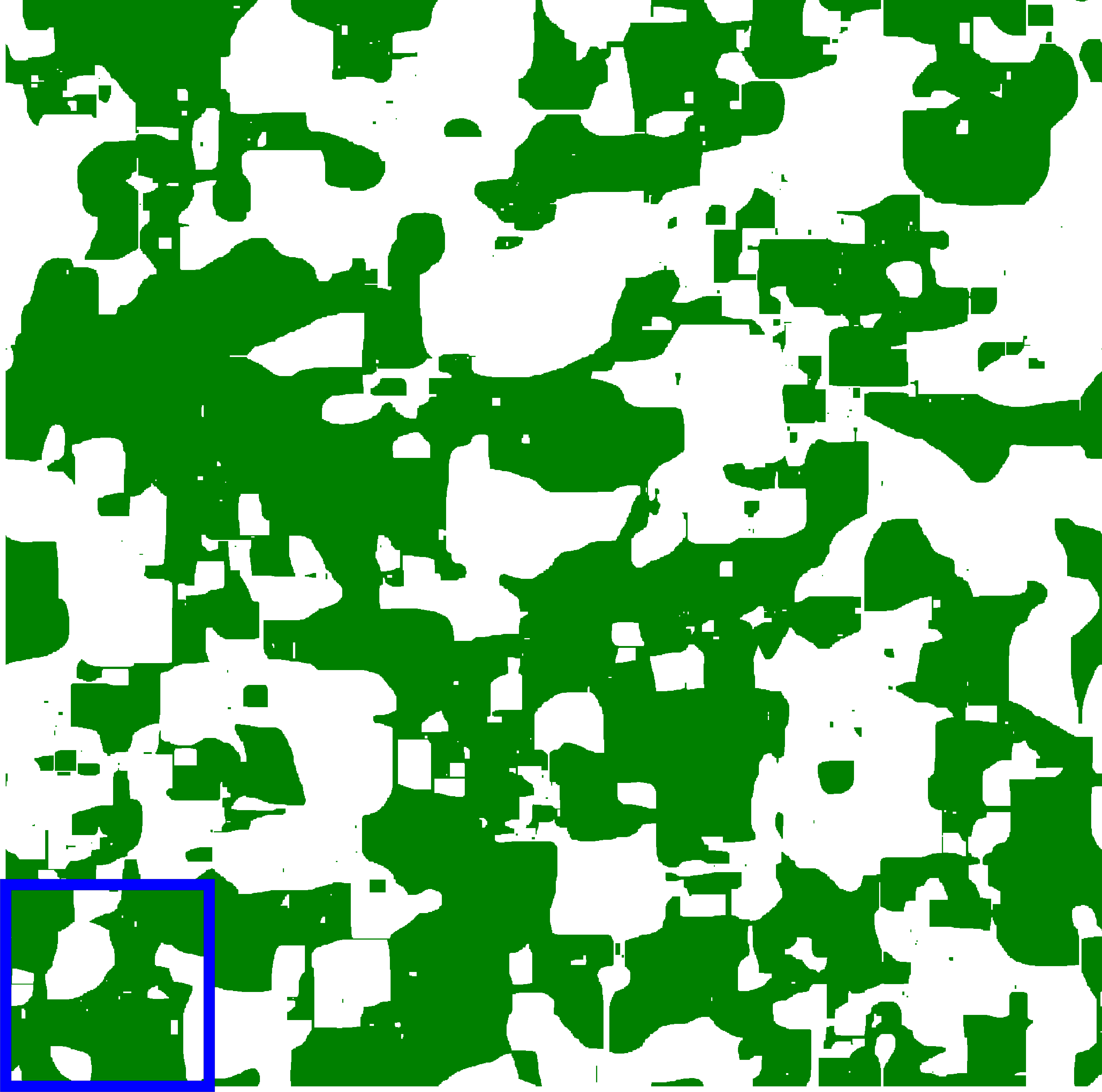} & 
    \includegraphics[width=0.15\textwidth]{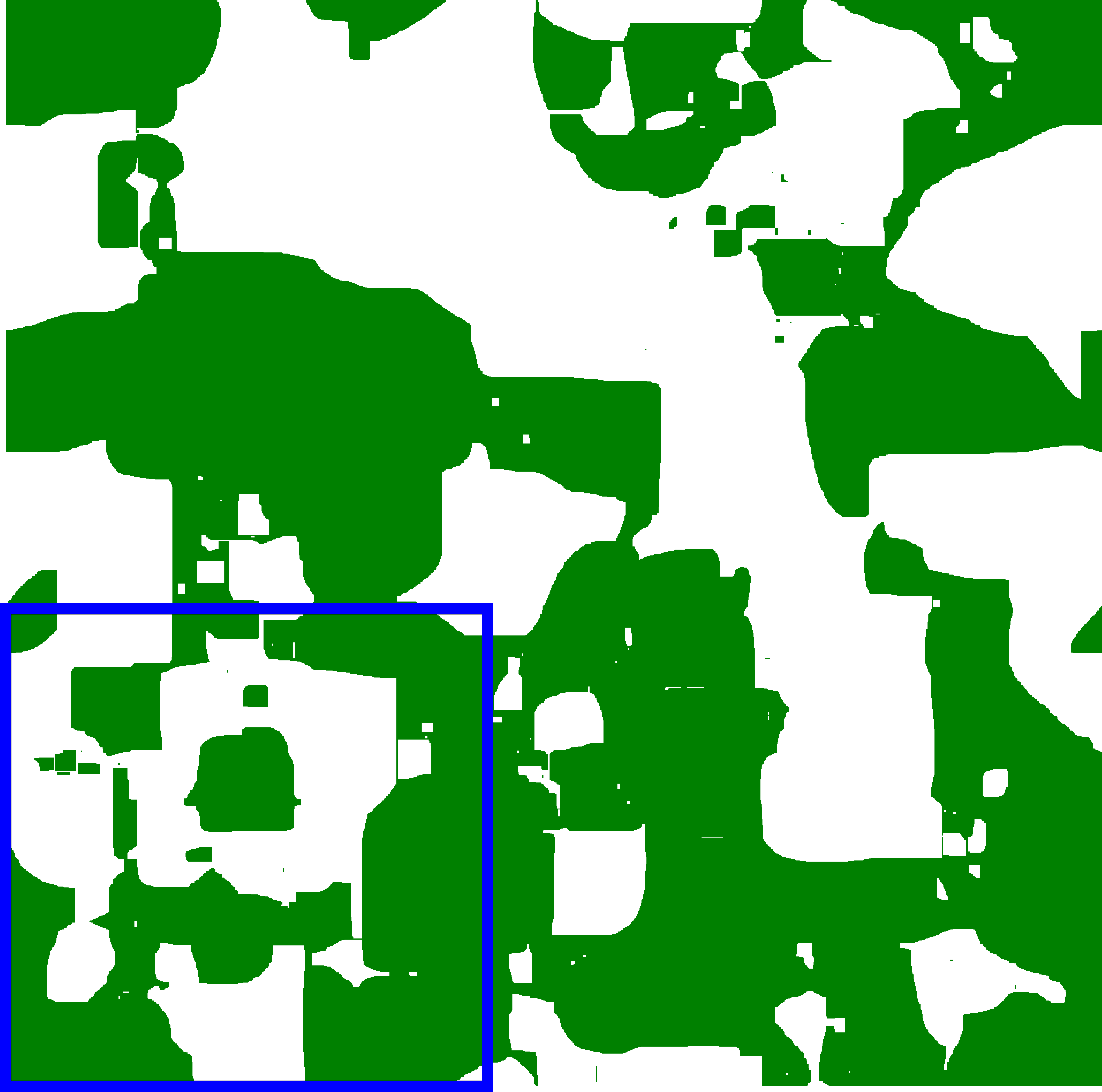} & 
    \includegraphics[width=0.15\textwidth]{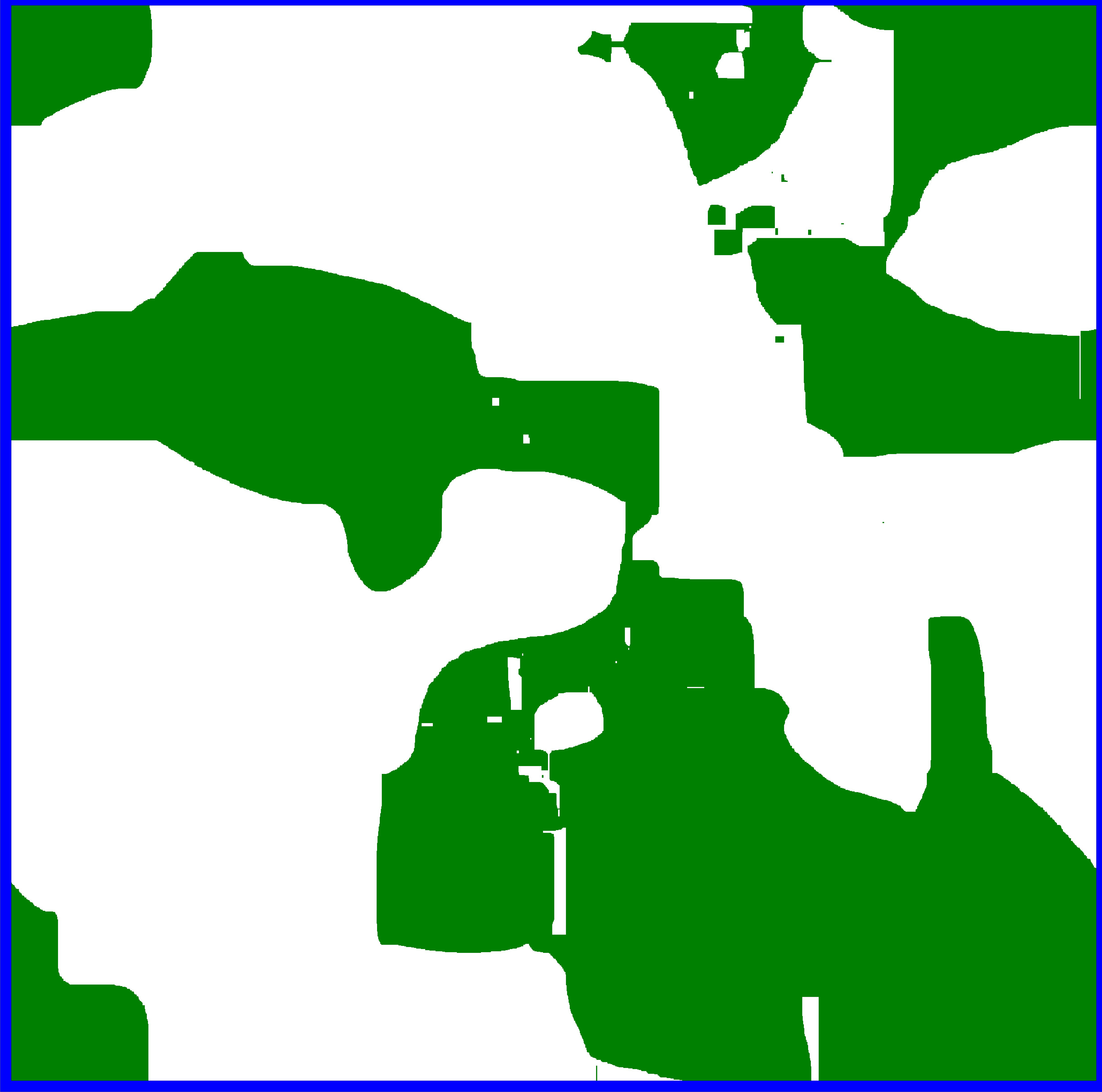}                                                             \\
    \includegraphics[width=0.15\textwidth]{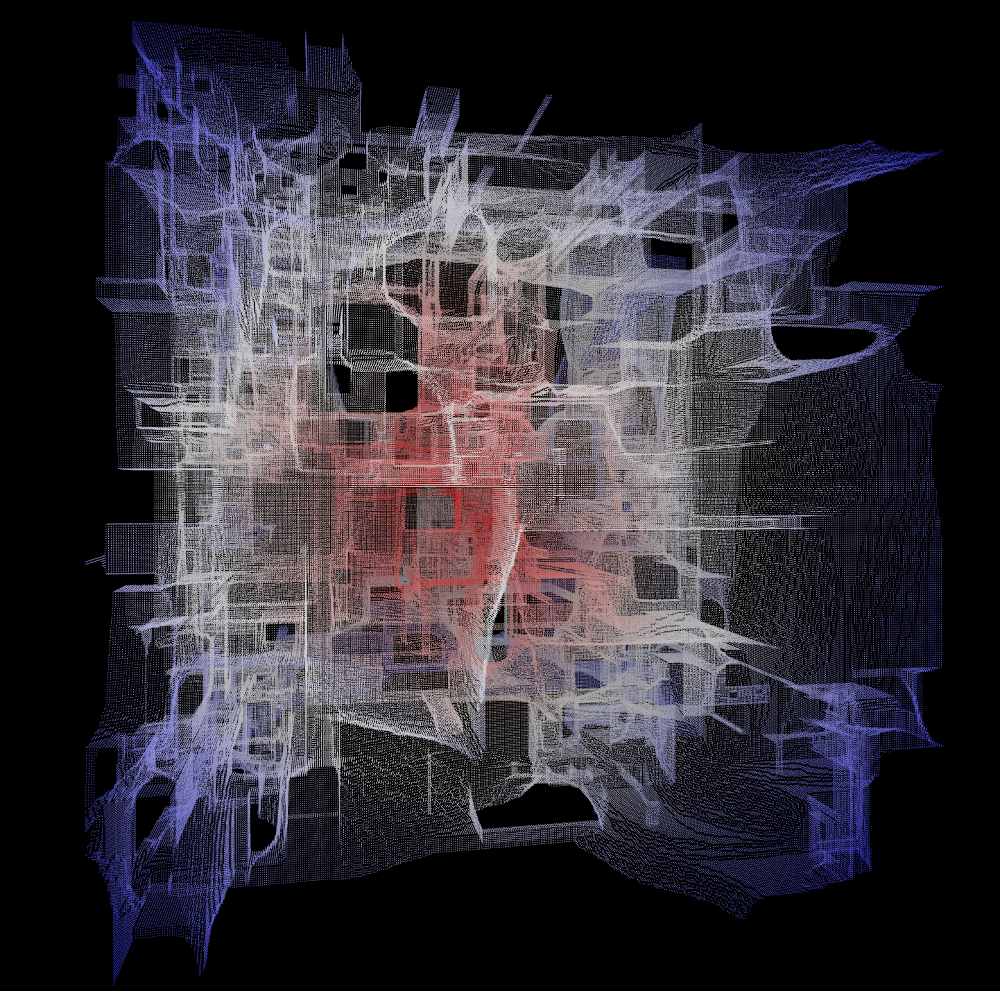}                      & 
    \includegraphics[width=0.15\textwidth]{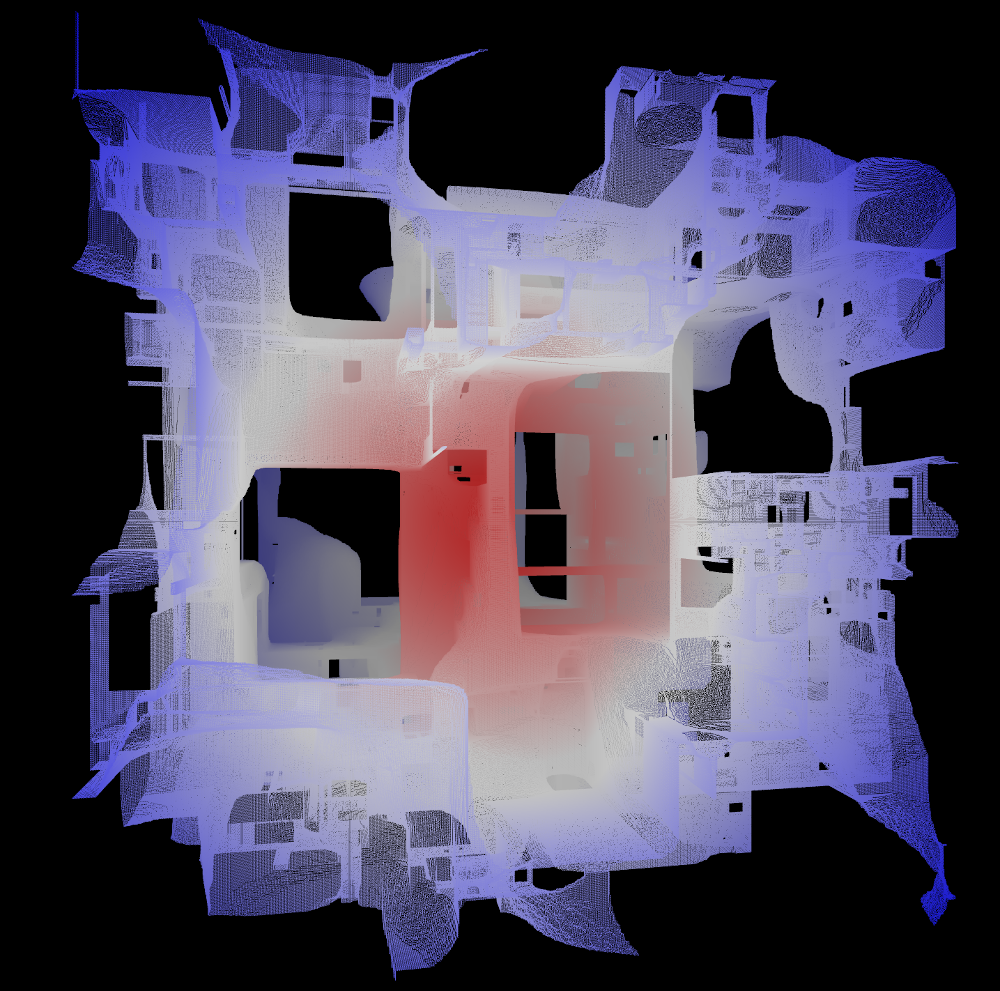}                      & 
    \includegraphics[width=0.15\textwidth]{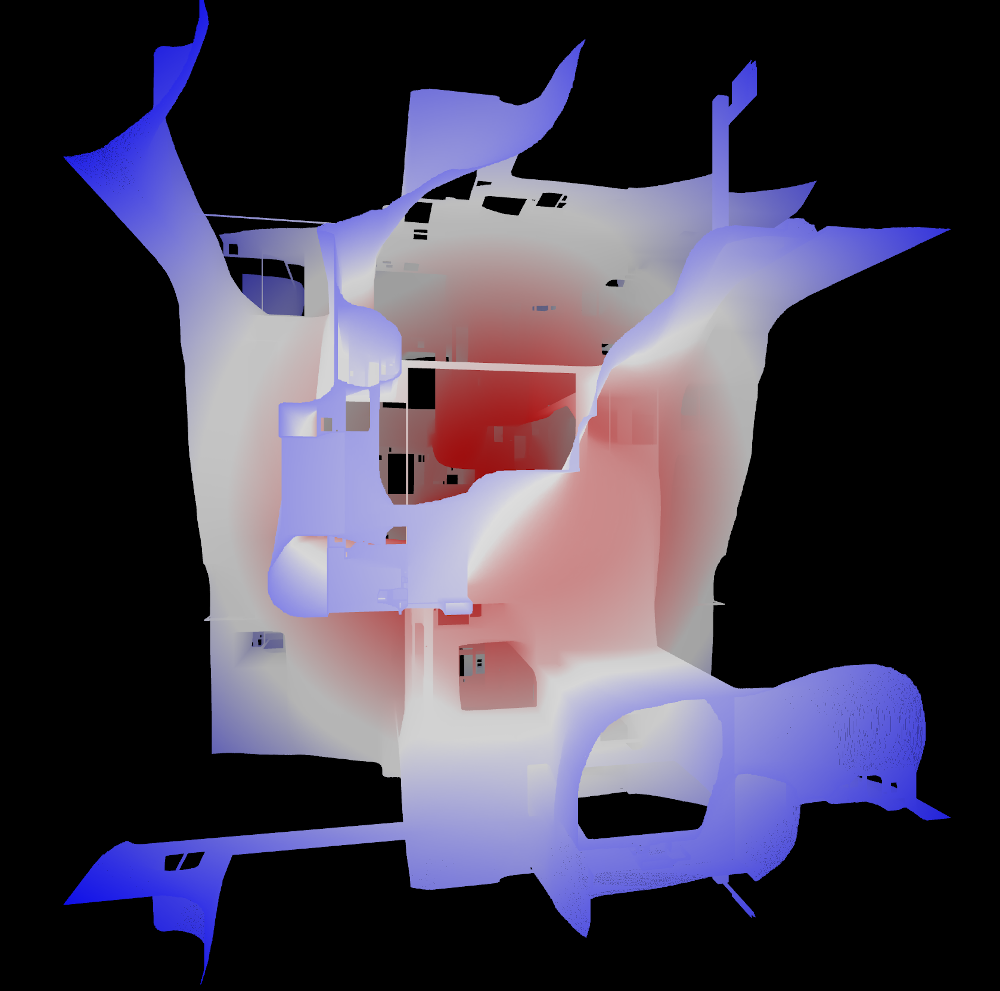}                                                                                  \\
    $t=10^4~\text{MCS}$                                                    & $t=6\times 10^4~\text{MCS}$ & $t=3 \times 10^5~\text{MCS}$
  \end{tabular}
  \caption{ \textbf{Visualization of the three-dimensional Ising configurations at different times of the quench.}
    Cross-sections (top) and three-dimensional snapshots (bottom) of the spin configuration for $L=2048$ at various times in units of Monte Carlo sweeps (MCS). The cross-sections in the top panel are cuts through the full lattice, where down-spins (the minority direction in this realization) are colored green. 
    The marked square subsections of linear extension $K \simeq 10\,\ell(t)$ are shown in the bottom panels as three-dimensional visualizations highlighting the interfaces between domains. The color (red-white-blue) indicates the distance from the center of the subsection. For details on the visualizations see Supplementary Section III and for snapshots at more times see Supplementary Fig.~5.
  }
  \label{figure:snapshots}
\end{figure}

Nonetheless, direct simulations of the continuous and deterministic time-dependent Ginzburg-Landau equation for big systems have provided the correct value of $\alpha=1/2$~\cite{brown2002numerical}.
In recent work~\cite{chakraborty2017coarsening,das2017kinetics,vadakkayil2019finite} the three-dimensional problem was tackled again by simulating this process using very big simple-cubic lattices with linear size up to $L=750$, from which the authors conjectured a crossover to $\alpha=1/2$ at late times.
\begin{figure*}
  \includegraphics{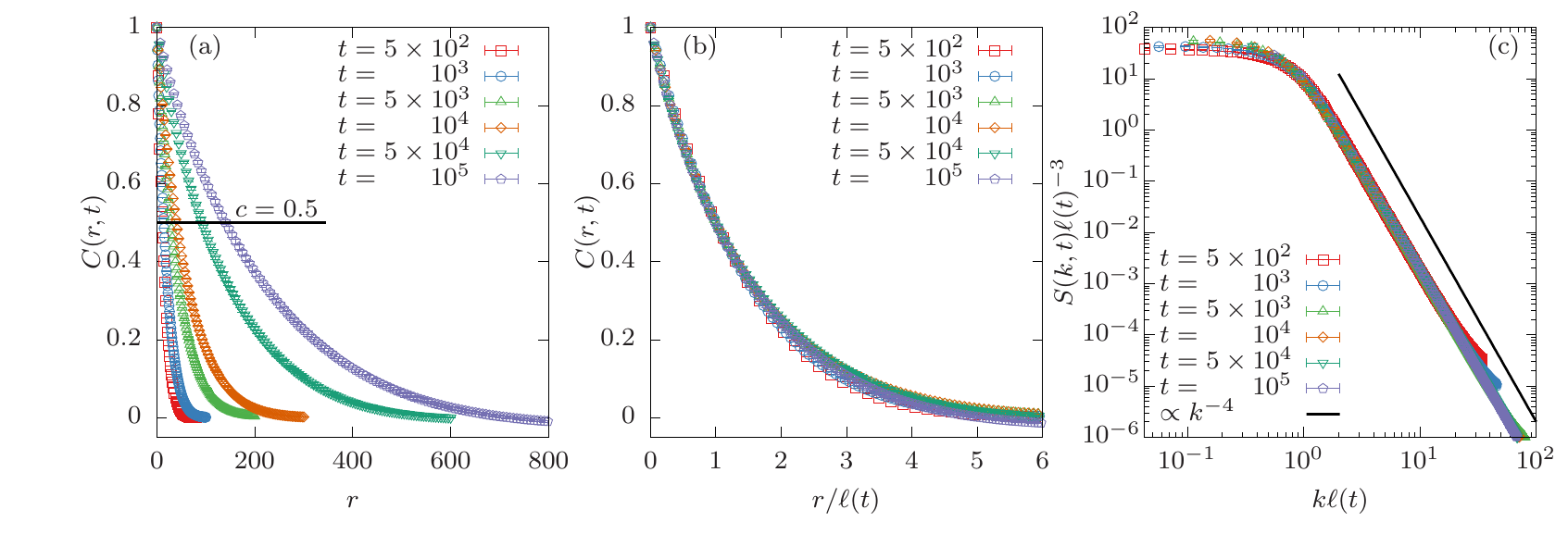}
  \caption{
    \textbf{Demonstration of scaling of the two-point correlation function and structure factor.}
    (a) Correlation function $C(r,t)$ versus distance $r$ for $L=2048$ and several times $t=500,\dots,10^5$.
    For increasingly later times, the correlation function decays slower, indicative of a growing length scale.
    (b) Showcase of self-similarity by plotting $C(r,t)$ against $r/\ell(t)$.
    (c) Structure factor $S(k,t)$ scaled to collapse, i.e., $S(k,t)\ell(t)^{-3}$ against $k\ell(t)$.
    The solid line is a power-law $\sim k^{-4}$, where $-4$ is the expected exponent of Porod's law. Error bars correspond to the standard error.
  }
  \label{figure:combined}
\end{figure*}

In an attempt to solve this long-standing puzzle, we performed Monte Carlo (MC) simulations of much larger systems with up to $L=2048$ (using periodic boundary conditions), corresponding to more than $8$ billion spins by employing a memory and time efficient GPU implementation.
Our implementation is adapted from a publicly available code~\cite{Barash2017} that uses a checkerboard decomposition of the system.
\par

\section{Results and Discussion}

In Fig.~\ref{figure:snapshots} we present visualizations of the lattice configuration of an exemplary simulation run for $L=2048$ at times $t=10^4$, $6\times 10^4$, and $3 \times 10^5$ in units of MC sweeps (MCS).
The top row shows plane cuts of the configuration allowing for an easy comparison with the well-known smooth behavior in $d=2$ (see, e.g., Fig.~2 in Ref.~\cite{bray2002theory}). 
Three-dimensional representations of domain interfaces are shown in the bottom panel.
For early times (left panels), one observes a roughening of the domain boundaries as reported several times earlier for zero-temperature quenches in $d=3$ spatial dimensions. 
This is clearly in violation of the arguments used to derive $\alpha=1/2$ where a diffusive domain-curvature minimization is assumed, so that here another effective growth exponent is to be expected.
Contrasting, at intermediate and even more so at late times (middle and right panels) the domains appear much smoother and diffusion-like growth might be anticipated.
However, as domains inside domains are a prominent feature in earlier snapshots but not as much at late times, during the coarsening process annihilation of these domains has to take place.
We conjecture that this annihilation is an additional contribution to the domain growth.
\par

To quantify these observations, we measure the two-point equal-time correlation function
\begin{equation}
  C(r,t) = \langle s_i s_j \rangle - \langle s_i\rangle \langle s_j\rangle, \label{eq:correlation function}
\end{equation}
where $\langle \cdot \rangle$ denotes the average over initial conditions and independent trajectories.
With increasing order of the system, one expects the correlation function to decay slower, i.e., for late times the correlation function should correspondingly indicate a stronger correlation.
Demonstration of this is shown in Fig.~\ref{figure:combined}(a) for the times mentioned in the key for $L=2048$ and $T=0$.
All data was obtained by starting from random spin configurations (with magnetization $m \approx 0$) and averaging over $40$ independent realizations (we use the same number of independent realizations for each system size).
Note, that previous work~\cite{Chakraborty2015} indicates that similar results are to be expected from finite starting temperatures.

$C(r,t)$ is expected to follow dynamical scaling, i.e.
\begin{equation}
  C(r,t) = \tilde{C} \left(r / \ell\left(t\right)\right).
\end{equation}
This is self-consistently tested by extracting $\ell(t)$ from the intersection of $C(r,t)$ with a constant value of $c=0.5$. (For the effect of different choices of $c$, see Supplementary Discussion III.)
Subsequently we plot $C(r,t)$ versus $r/\ell(t)$ in Fig.~\ref{figure:combined}(b).
We note that especially at large distances $r$ the data collapse is not optimal.
This may be an indication of a number of things, e.g., the occurrence of finite-size effects.
Another indicative explanation is that the growth exponent $\alpha$ is not yet a constant but effectively varies with $t$.
\par
Additionally, one looks at the structure factor $S(k,t)$ which is the Fourier transform of the correlation function.
This quantity, similarly to the correlation function, collapses when properly rescaled, i.e., by plotting $S(k,t)\ell(t)^{-3}$ versus $k\ell(t)$ as shown in Fig.~\ref{figure:combined}(c). The data for different sizes collapses quite well and shows a clear power-law decay with Porod's exponent $d+1=4$.

\begin{figure}
  \includegraphics{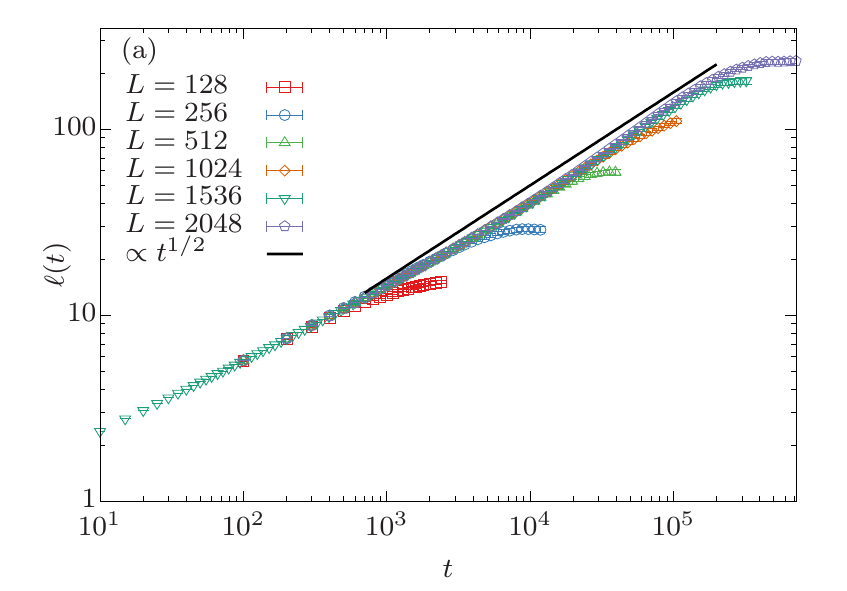}
  \includegraphics{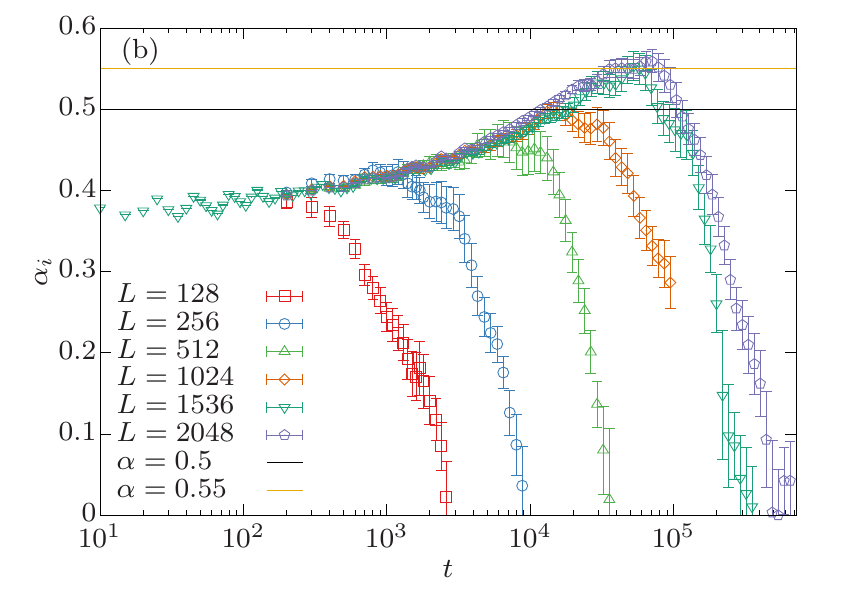}
  \caption{
    \textbf{Length scale as a function of time and instantaneous growth exponent.}
    (a) Length scale $\ell(t)$ for quenches to $T=0$ in $d=3$ spatial dimensions with linear size $L=128,\dots,2048$ on a log-log scale.
    The black solid line shows the expected power-law behavior of $\ell(t) \sim t^{1/2}$.
    (b) Instantaneous exponent $\alpha_i$ is shown against $t$ for the same data.
    Here the $x$-axis is logarithmic to highlight the large $t$ behavior where $\alpha_i$ takes values consistently larger than~$1/2$. Error bars correspond to the standard error.
  }
  \label{figure:ell}
\end{figure}

Finally, we present the characteristic length $\ell(t)$ versus $t$ on a log-log scale in Fig.~\ref{figure:ell}(a) for system sizes $L=128,256,512,1024,1536,2048$. Initially, $\ell(t)$ grows independently of the system size $L$ as individual domains can grow unrestrictedly. It is only when $\ell(t)$ reaches a value of the order of magnitude of $L$ that domain growth is hindered by the finite nature of the lattice which shows itself in the form of finite-size effects that end in a stagnation of the growth for that system size.
The solid line shows the asymptotic growth law $\ell(t) \sim t^{1/2}$ as predicted, where one clearly sees that this is not parallel to the data for late times.
To get a more detailed impression of this, we show in Fig.~\ref{figure:ell}(b) the instantaneous exponent
\begin{equation}
  \alpha_i(t) = \frac{d \ln \ell(t)}{d \ln t},
\end{equation}
that is, the local slopes in the top panel of Fig.~\ref{figure:ell}(a).
At very early times $\ell(t)$ grows like $t^{\alpha_i}$ with $\alpha_i$ compatible with $0.35-0.40$. When only considering lattice sizes up to $L=256$, then only this behavior can be seen as was the case in Refs.~\cite{shore1992logarithmically,cueille1997spin,lipowski1999anomalous}. Reference~\cite{Corberi2008} observed $t^{0.43}$ using a lattice size of $L=512$, which is in good agreement with our measurement for this size. For $L=1024$ we observe $\alpha_i \approx 1 / 2$ for a short time, and for $L=1536$ and $L=2048$ we observe an exponent $\alpha_i > 1/2$, which is completely unexpected from existing simulations and theory. As the two largest system sizes show the $\alpha_i(t) > 1/2$ signal before the onset of finite-size effects in either system size, we conclude that this signal should persist at these times ( i.e., $t \in [2\times 10^4,7\times 10^4]$ ) as ${L\rightarrow\infty}$.
We conjecture that the aforementioned contribution to the domain growth from the annihilation of domains inside domains may be the cause of this superdiffusive behavior with $\alpha_i > 1/2$.

To assure that this behavior is not an artifact from the concurrent spin update caused by the checkerboard decomposition of the system, we repeated our measurements of the domain size $\ell(t)$ with a number of different update algorithms including an efficient $n$-fold way simulation~\cite{n-fold} for system sizes up to $L=1536$ and found agreement within error bars; see Supplementary Discussion II for a comparison of the results from the algorithms and Supplementary Methods I for a discussion of their implementations.

\par
By studying significantly larger system sizes than available in the literature, we thus discover yet another twist in the coarsening story of the three-dimensional Ising model at zero temperature.
We find strong evidence for $\alpha_i(t)$ at least pre-asymptotically taking values significantly larger than $1/2$ which is in conflict with previous numerical conjectures that $\alpha = 1 / 2$ using smaller systems~\cite{das2017kinetics,vadakkayil2019finite}; thus again challenging our understanding of the dynamics in this simple model.
(The maximal value obtained for $\alpha_i$ exceeds $1/2$ by four [three] times the standard error for $L=2048$ [$L=1536$].)
The structure of the domains has been described as sponge-like~\cite{olejarz2011zero} or fractal~\cite{Corberi2008}.
Anomalous diffusion, including both sub- and superdiffusion, is a well known phenomenon on fractal structures~\cite{Metzler2000}.
Hence, we believe that the peculiar structure of domains found in this coarsening problem is both the reason for the early time behavior with $\alpha_i<1/2$ and the late-time stage with $\alpha_i>1/2$.
It is nonetheless possible, that we recover $\alpha=1/2$ in the thermodynamic limit, that is in the double limit of $L\rightarrow\infty$ and $t\rightarrow\infty$.

\begin{figure*}
  \includegraphics[width=0.3\textwidth]{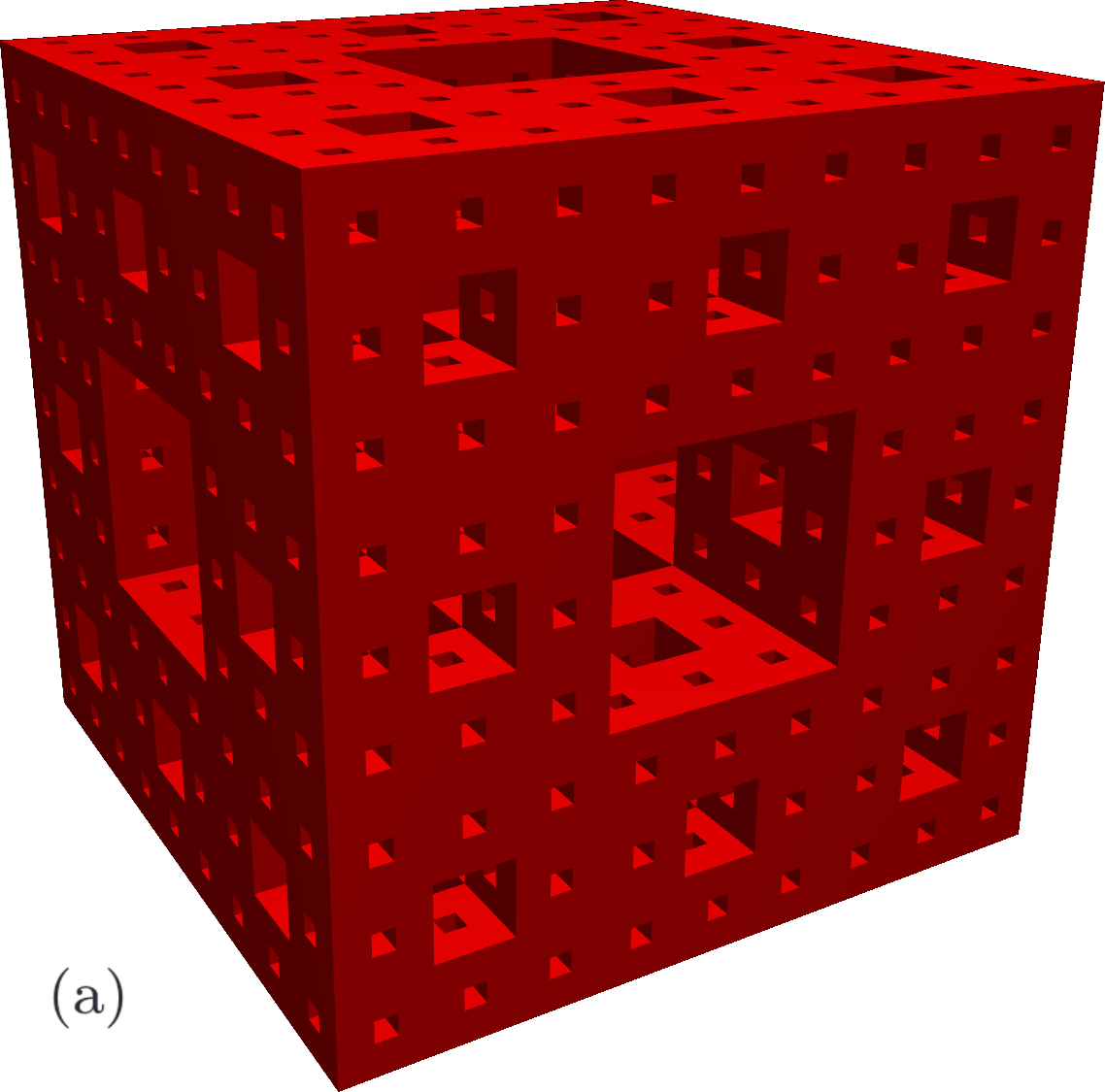}\hfil
  \includegraphics[width=0.3\textwidth]{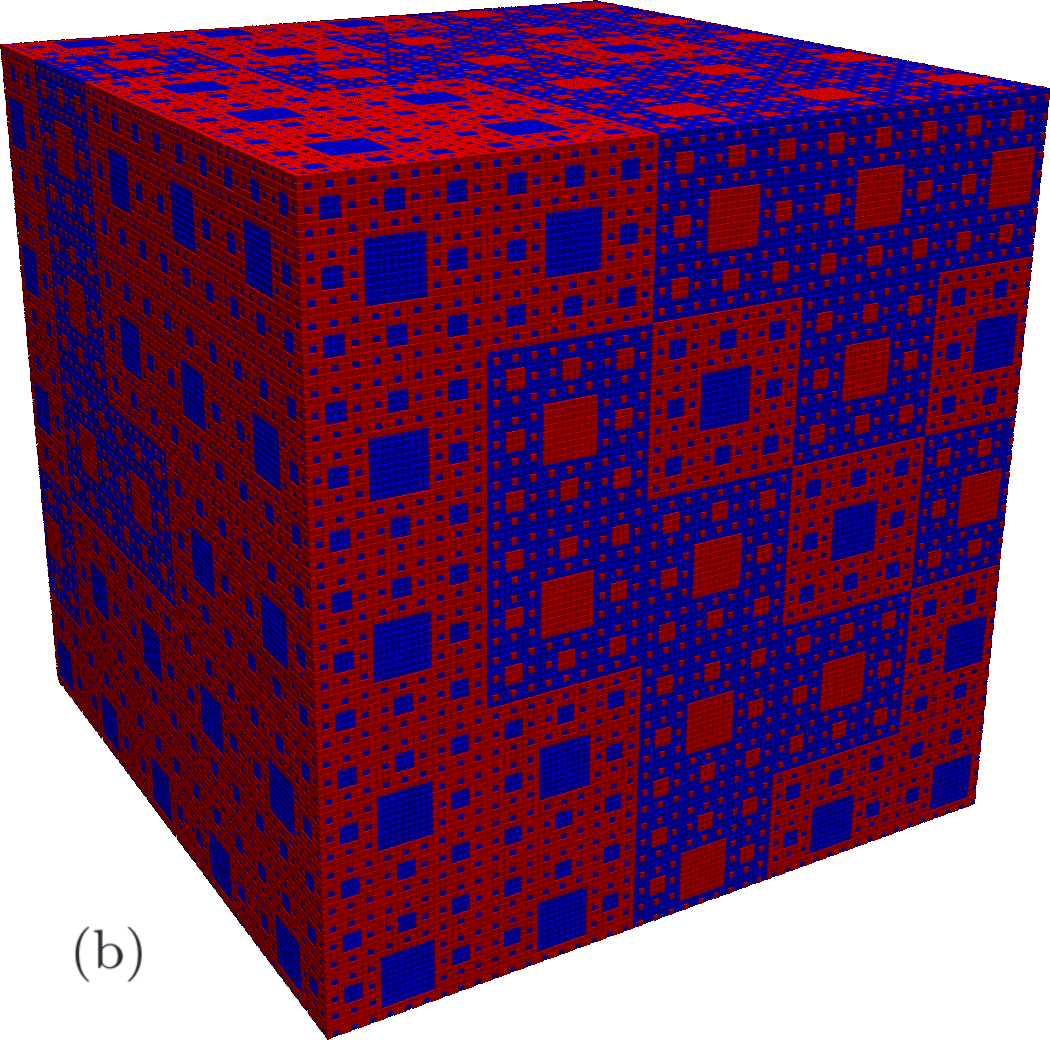}\hfil\hspace*{-1cm}
  \hspace{2mm}\hfil\includegraphics{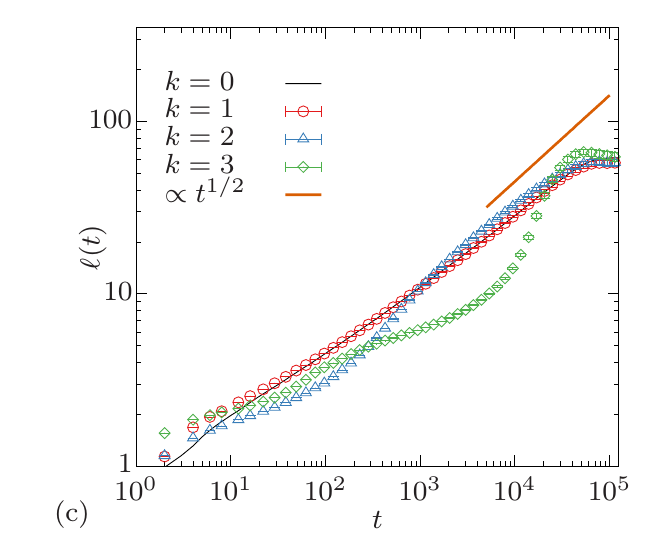}\hspace{-5mm}
  \caption{
    \textbf{Results for zero-temperature coarsening using an artificial sponge structure as starting configuration.}
    (a) Third iteration ($k=3$) Menger sponge of size $27^3$. (b) $128^3$ initial Ising configuration consisting of third iteration Menger sponges, red (blue) corresponding to up (down) spins. (c) Length scale $\ell(t)$ of a quench to $T=0$ starting from such an artificial configuration using Menger sponges of $k^\text{th}$ iteration and using $L=512$. Error bars correspond to the standard error.
  }
  \label{figure:sierpinski}
\end{figure*}

To test our intuition that the sponge-like behavior is responsible for this superdiffusive growth, we carry out one further test. We replace the initial high-temperature random configuration by an artificial sponge structure to probe its effect on the dynamics. As a prototypical sponge structure we use Menger sponges~\cite{Mandelbrot1977}, the three-dimensional generalization of Sierpinski carpets [see Fig.~\ref{figure:sierpinski}(a)]. The starting configuration [see Fig.~\ref{figure:sierpinski}(b)] of dimensions $L^3$ is created by repeating sponges of $k^\text{th}$ iteration of size $(3^k)^3$. For each sponge we pick uniformly at random whether the Menger structure is represented by up or down spins.

We carry out a zero-temperature quench on these structures in the same manner as before but using the $n$-fold way update~\cite{n-fold} (see Supplementary Methods I C for more detail) instead as this choice avoids potential interference of the Menger sponge structure with the structure of the checkerboard decomposition.
From this we obtain the characteristic length scale $\ell(t)$ presented in Fig.~\ref{figure:sierpinski}(c). Clearly, the significant differences between $k=0$ (corresponding to our case from before, i.e., a quench from $T=\infty$) and higher-order fractals with $k\neq 0$ become more pronounced the larger $k$. We note two key effects: On the one hand, at early times the dynamics for $k\neq 0$ is much slower than in the $k=0$ case and on the other hand, at later times it becomes much faster than the original dynamics and clearly exceeds a growth governed by $\propto t^{1/2}$. From this we learn that indeed sponge-like structures can cause anomalously slow early dynamics followed by superdiffusive growth at later times which is reminiscent of our observation when quenching the three-dimensional Ising model from infinite to zero temperature. 

To conclude, we have simulated zero-temperature coarsening of the three-dimensional Ising model with nearest-neighbour interactions.
For this model, the growth exponent of the characteristic length scale is predicted to be $1/2$, whereas most simulations previously suggested a smaller exponent $\approx 1/3$.
Using a highly efficient GPU implementation, we simulate this process and are able to go to linear system sizes of $L=2048$, i.e., over $8$ billion spins. This allows us to monitor late times which previously were not accessible and we discover a previously unknown superdiffusive growth behavior which we attribute to the annihilation of sponge-like structures emerging at early times.

Although we expect $1/2$ for the growth exponent in the long-time limit, we cannot fully verify this expectation. This is due to the presence of pre-asymptotic effects at late times even for very large systems. Based on preliminary investigations (not presented here) we are confident that we may get access to the necessary hardware to study even larger systems, i.e., $L = 4096$, in the near future. Additionally, very recent work~\cite{Vadakkayil2022} reported on the anomalously slow growth prevailing even for quenches to $T>0$ as long as the quench temperature is well below the roughening transition temperature $T_R$. We will investigate in future work whether also the superdiffusion-like behavior is seen at these temperatures.
\par
The idea of making use of GPUs for nonequilibrium investigations using MC simulations is expected to spark investigations of bigger systems also for related spin models.
While GPU implementations of MC simulations of spin models have been used and studied in the equilibrium context for several years \cite{tomov2005benchmarking,preis2009gpu,block2010multi,weigel2011simulating,weigel2011gpu,weigel2012performance,weigel2012simulating,yavors2012optimized}, the potential of application of this approach to nonequilibrium simulations has not been fully realized.
One possible explanation for this neglect in nonequilibrium studies is the necessary reliance on checkerboard decomposition to speed up the simulations on the GPU, which some may suspect to introduce artifacts in the dynamics of the simulation.
However, with our work we demonstrate that this fear appears to be unfounded and various GPU update methods are indeed suitable for nonequilibrium studies.

\section{Methods}
\subsection{Spin updates}
In the following we discuss the checkerboard update as an alternative to the random-site-flip update. For details on further alternative update methods, we refer to Supplementary Methods I.
\subsubsection{Random-site-flip update}
The random-site-flip (rsf) update is the most straightforward method to perform MC simulations and to study coarsening in the Ising model.
In each MC step one chooses a site $i$ at random and proposes to flip the spin $\sigma_i \in \{-1, +1\}$. Based on the change in energy $\Delta E$ attributed to the proposed move, in general for non-zero temperature $T$ it is accepted with the Glauber acceptance probability~\cite{glauber1963time}

\begin{equation}
  p_\text{acc}(\Delta E,T) = \frac{1}{1+e^{\Delta E / k_B T}},
\end{equation}
where the Boltzmann constant $k_B$ usually is set to unity to fix units. In the limit $T\rightarrow 0$ this simplifies to
\begin{equation}
  p_\text{acc}(\Delta E) = \begin{cases} 0, &\, \Delta E > 0 \\ \frac 1 2, &\, \Delta E = 0 \\ 1, &\, \Delta E < 0 \end{cases}, \label{equ:glauberAcceptance}
\end{equation}
which is the acceptance probability we use throughout.
$N=L^3$ such MC steps are referred to as one MC sweep (MCS), where $L$ is the linear lattice size.

Clearly, this approach is rather inefficient as \emph{i)} a significant amount of computing resources is wasted on proposing moves with $\Delta E > 0$~\cite{olejarz2011zero}, which always are rejected, and \emph{ii)} because it is inherently sequential 
making it hard to parallelize the algorithm.

\subsubsection{Checkerboard update}
The key idea of many domain-decomposition spin update algorithms such as the checkerboard update is that with local, i.e., short-range, interactions only, the lattice can be decomposed into sub-lattices such that spins of the same group do not interact with one another. 
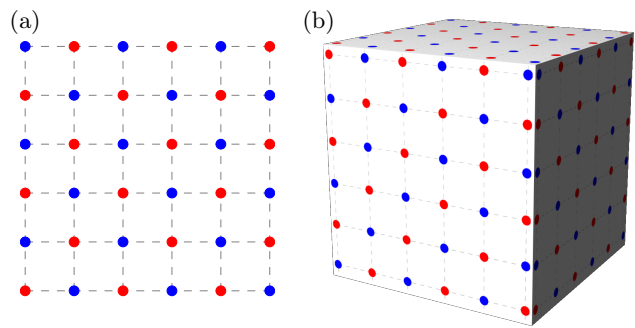
\begin{figure}[b]
  \centering
  \begin{tikzpicture}[scale=0.65]
    \node at (0,5.5) {(a)};
    \node at (6,5.5) {(b)};
    \clip (-0.1,-0.7) rectangle (5.1cm,5.3cm);
    \coordinate (Bone) at (0,2);
    \coordinate (Btwo) at (2,-2);
    \draw[style=help lines,dashed] (0,0) grid[step=1cm] (5,5);
    \foreach \x in {0,...,5}{
        \foreach \y in {0,...,5}{
            \pgfmathsetmacro{\count}{\x+\y}
            \ifodd \count
              \node[draw,circle,inner sep=1.3pt,fill,blue] at (\x,\y) {};
            \else
              \node[draw,circle,inner sep=1.3pt,fill,red] at (\x,\y) {};        
            \fi
          }
      }
  \end{tikzpicture}
  \hfil
  {\hspace*{-0.6cm}\vspace{-0.2cm} \includegraphics[width=0.23\textwidth]{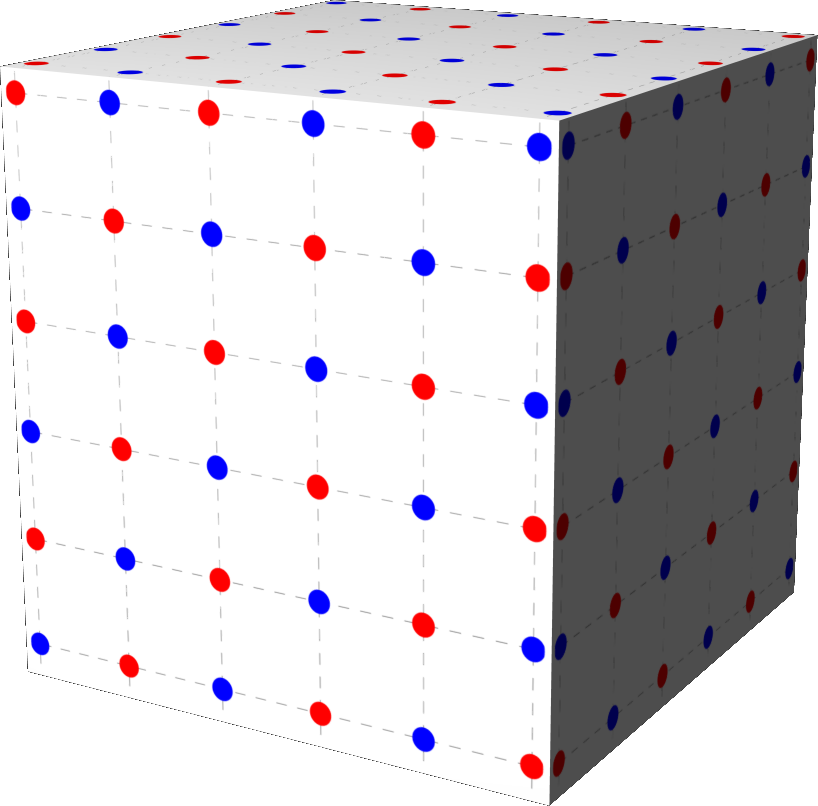}}
  \caption{\textbf{Checkerboard decomposition.} All red (blue) sites can be updated simultaneously as they only depend on blue (red) sites. (a) Checkerboard decomposition in two spatial dimensions. (b) Generalization to three spatial dimensions.}
  \label{fig:checkerboard}
\end{figure}

In the case of the two-dimensional square lattice with only nearest-neighbor interactions one of the simplest such decompositions looks like a checkerboard (see Fig.~\ref{fig:checkerboard}), hence the name of the method. One MC sweep then consists of \emph{i)} updating all red spins concurrently with $N/2$ parallel threads, followed by \emph{ii)} updating all blue spins concurrently with $N/2$ parallel threads. Equivalently, one may choose to update all blue spins first (see Supplementary Methods I B for more detail). Each proposed spin flip is accepted with the same probability as before [see Eq.~(\ref{equ:glauberAcceptance})] and the only difference as compared to the rsf update is the order in which updates are proposed.
The generalization to $d=3$ is conceptually straightforward, see the right panel of Fig.~\ref{fig:checkerboard}.

This update scheme is particularly suited for an implementation on graphics processing units (GPUs). 
GPUs have several thousand threads which can be used to update the independent spins in parallel. 
Our implementation in CUDA for this update scheme is based on the code from Ref.~\cite{Barash2017} although heavily adapted as the authors optimized their code for $>1000$ simultaneous simulations of small systems on a single GPU. In contrast, we simulate a single large system per GPU. Additionally, Ref.~\cite{Barash2017} considers the two-dimensional Ising model. Hence, the respective parts of the code have been modified accordingly.

\subsection{Calculating the correlation function}
Na\"ive calculation of the correlation function defined in Eq.~(\ref{eq:correlation function}), i.e.,
\begin{equation}
  C(r,t) = \langle s_i s_j \rangle - \langle s_i\rangle \langle s_j\rangle, \tag{\ref{eq:correlation function}}
\end{equation}
involves a double summation over all spins requiring $\mathcal{O}(N^2)$ time.
Using a Fast Fourier Transform (FFT) allows the calculation of $\langle s_i s_{i+k} \rangle$ in $\mathcal{O}(N \log N)$, i.e.,
\begin{equation}
  \overline{s_i s_{i+k}} = [\mathcal{F}^{-1}(|\mathcal{F} s_i|^2)]_k, \label{eq:sisik}
\end{equation}
where $\mathcal{F}$ is the three-dimensional discrete Fourier transformation operator and the overline stands for an average over $i$, exploiting the translational invariance.
$C(r,t)$ shown in the main text is then obtained by radially averaging over the three-dimensional correlation matrix.

In standard FFT routines two \texttt{double} values are used per (spin) site both for input and output, requiring thus $4\times 8$ bytes per spin. For $L=2048$ this amounts to $4\times 8 \times 2048^3= 2^{38} $ bytes $= 256$ GB RAM necessary to carry out the FFT which on modern CPU computing nodes is possible but still quite restrictive as it limits the number of simulations which can be run in parallel on the same node. We use the FFTW library~\cite{frigo2005design} which allows for in-place calculation such that the memory footprint is cut in half.

Further, spin variables only take the values $-1$ and $+1$, and $|\mathcal{F} s_i|^2$  in Eq.~(\ref{eq:sisik}) only real values. Hence, the real-data discrete Fourier transform (DFT) routine can be used for both transforms which reduces the used memory by another factor of two by using that the resulting DFTs are Hermitian. This allows the input to be stored in $N$ \texttt{double} values and the output to be stored in $N/2$ complex (= two \texttt{double}s) values. When using in-place real-data DFT, this brings the memory footprint down to eight bytes per site such that for the FFT of one $2048^3$ lattice only $64$ GB RAM are necessary which are readily available on our compute nodes. However, already for the next bigger system size, i.e., $L=4096$, we can no longer compute the FFT in the same manner as $64 \times 8 = 512 \text{ GB}$ RAM are not available to us. 

Additionally, FFTW supports multi-threading such that we can speed up the calculation by a factor of about $10$ compared to the sequential algorithm.

\section{Data availability}
The data that support the findings of this study are available from the corresponding author upon reasonable request.

%

\section{Acknowledgements}

We thank Martin Weigel for helpful discussions. This project was funded by the Deutsche Forschungsgemeinschaft (DFG, German Research Foundation) under project Nos. JA 483/33-1 and 189\,853\,844 -- SFB/TRR 102 (project B04), and the Deutsch-Franz\"osische Hochschule (DFH-UFA) through the Doctoral College ``$\mathbb{L}^4$'' under Grant No.\ CDFA-02-07. We further acknowledge support by the Leipzig Graduate School of Natural Sciences ``BuildMoNa''.

GPU computations for this work were done using resources of the Leipzig University Computing Centre. The $n$-fold way calculations were performed using the Sulis Tier 2 HPC platform hosted by the Scientific Computing Research Technology Platform at the University of Warwick. Sulis is funded by EPSRC Grant EP/T022108/1 and the HPC Midlands+ consortium. Moreover, we acknowledge the provision of computing time on the parallel computer cluster \emph{Zeus} at Coventry University.

\section{Author contributions}
H.C. and W.J. conceived the work; D.G. performed the numerical simulations; D.G. and H.C. analyzed the data; all authors discussed the results and wrote the manuscript; W.J. supervised the work.

\section{Competing interests}
The authors declare no competing interests.

\end{document}


\title{Supplementary Material: Superdiffusion-like behavior in zero-temperature coarsening of the $d=3$ Ising model}
\author{Denis Gessert}
\email{denis.gessert@itp.uni-leipzig.de}
\affiliation{Institut f\"ur Theoretische Physik, Universit\"at Leipzig, IPF 231101, 04081 Leipzig, Germany}
\affiliation{Centre for Fluid and Complex Systems, Coventry University, Coventry CV1~5FB, United Kingdom}
\author{Henrik Christiansen}
\email{henrik.christiansen@itp.uni-leipzig.de}
\altaffiliation{Present address: NEC Laboratories Europe GmbH, Kurfürsten-Anlage 36, 69115 Heidelberg, Germany.}
\affiliation{Institut f\"ur Theoretische Physik, Universit\"at Leipzig, IPF 231101, 04081 Leipzig, Germany}
\author{Wolfhard Janke}
\email{wolfhard.janke@itp.uni-leipzig.de}
\affiliation{Institut f\"ur Theoretische Physik, Universit\"at Leipzig, IPF 231101, 04081 Leipzig, Germany}
\date{\today}

\maketitle

\section{Supplementary methods: Alternative spin updates}
We study coarsening, a nonequilibrium process, in the three-dimensional Ising model with non-conserved order parameter by means of Monte Carlo (MC) simulations. 
As it is not uniquely specified how to perform spin updates, there are several ways to realize the underlying Markov chain.
In the case of nonequilibrium investigations clearly one cannot take advantage of non-local updates such as the Wolff cluster algorithm.
Nonetheless, for spin models and local updates, the way in which the single-spin updates are proposed is not stipulated.
Instead, one has a choice to adapt this, e.g., to speed up simulations.
In the following we present different approaches to implement spin updates alternative to the random-site-flip update and checkerboard update which are discussed in the main text.
At the end of this section we will check numerically that all the considered approaches produce dynamics that are compatible (within error bars) up to a rescaling of time by a constant factor.

\subsection{Double-checkerboard update}
\begin{figure}
  \begin{tikzpicture}[scale=0.4]
    \clip (0.5,0.5) rectangle (16.5cm,16.5cm);
    \draw[draw=black,fill=gray!30] (4.5,7.5) rectangle ++(4,1);
    \draw[draw=black,fill=gray!30] (4.5,12.5) rectangle ++(4,1);
    \draw[draw=black,fill=gray!30] (3.5,8.5) rectangle ++(1,4);
    \draw[draw=black,fill=gray!30] (8.5,8.5) rectangle ++(1,4);
    \coordinate (Bone) at (0,2);
    \coordinate (Btwo) at (2,-2);
    \draw[style=help lines,dashed] (-12,-12) grid[step=1cm] (20,20);
    \foreach \x in {1,...,16}{
      \foreach \y in {1,...,16}{
        \pgfmathsetmacro{\count}{floor((\x-1)/4)+floor((\y-1)/4)}	
        \ifodd \count
          \pgfmathsetmacro{\count}{\x+\y}
          \ifodd \count 
            \node[draw,circle,inner sep=1.5pt,fill,blue] at (\x,\y) {};
          \else
          \node[draw,circle,inner sep=1.5pt,fill,blue!50] at (\x,\y) {};        	
          \fi
        \else
          \pgfmathsetmacro{\count}{\x+\y}
          \ifodd \count 
            \node[draw,circle,inner sep=1.5pt,fill,red] at (\x,\y) {};
          \else
          \node[draw,circle,inner sep=1.5pt,fill,red!50] at (\x,\y) {};        	
          \fi			
        \fi
      }
    }
  \end{tikzpicture}
  \caption{Example double-checkerboard decomposition in ${d=2}$.}
  \label{fig:doubleCheckerboard}
\end{figure}
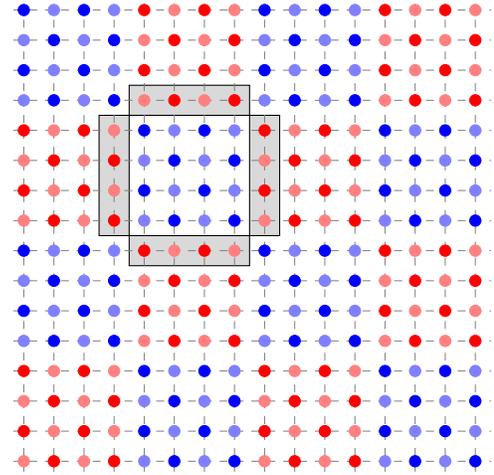

Weigel~\cite{[][{; the source code can be found at the author's personal homepage on \url{http://www.martin-weigel.org/research/gpu-computing}.}]weigel2011simulating} introduced a slightly more involved domain-decomposition scheme which aims at being more efficient on GPUs than the standard checkerboard approach. On GPUs, threads are organized in so-called thread blocks that can synchronize locally whereas threads from different blocks essentially cannot communicate within one execution of the GPU kernel.
For the performance of the algorithm it is beneficial to use the much faster shared memory of these blocks rather than the global memory available to all blocks. However, the shared memory is too small to even fit systems of moderate size. Thus, the idea of the double-checkerboard decomposition is to divide the system up into a coarse checkerboard (red and blue blocks in Fig.~\ref{fig:doubleCheckerboard}) of small sub-systems which still can fit in the shared memory. These can then be updated by a thread block using the standard checkerboard (lighter and darker sites in Fig.~\ref{fig:doubleCheckerboard}).

One MC sweep using the double-checkerboard scheme based on the decomposition depicted in Fig.~\ref{fig:doubleCheckerboard} then consists of the following steps:
\begin{enumerate}
  \item Start GPU kernel on red sublattice.
  \item Threads load spins into shared memory.
  \item Glauber update on darker spins.
  \item Barrier synchronization within thread blocks.
  \item Glauber update on lighter spins.
  \item Barrier synchronization within thread blocks.
  \item Writing changes to global memory.
  \item Start GPU kernel on blue sublattice and repeat steps 2 to 7.
\end{enumerate}

Our implementation is mostly based on the publicly available code of Ref.~\cite{weigel2011simulating}. Similar to checkerboard discussed in the main text, the generalization to three spatial dimensions was performed.

\subsection{Variants of the single- and double-checkerboard update}
\label{sec:variantsCheckerboard}
A priori it is not immediately clear that any of the described domain decomposition updates should exhibit the same kind of coarsening dynamics as the rsf update. In fact, there are at least three problems we are aware of regarding this type of update: \emph{i)} In contrast to rsf, no spin is proposed to be flipped more than once within a sweep, \emph{ii)} the violation of detailed balance, and \emph{iii)} extreme non-ergodicity at infinite temperature.

Regarding problem \emph{i)}, using the rsf update the chance for every spin in a system of $N$ spins to be proposed to be flipped exactly once during a sweep is $N! / N^N$ which approaches $\sqrt{2 \pi N} e^{-N}$ for large $N$. For a $2^3$ system this is approximately $0.2\%$ and already for $3^3$ around $2 \times 10^{-11}$. Thus, for the system sizes we study, it is nearly impossible not to have a repetition. Through generalization of the birthday problem~\cite{Mathis1991} we expect a repetition after $\mathcal{O}(\sqrt{N})$ proposals, that is $\mathcal{O}(\sqrt{N})$ repetitions per sweep.

Problem \emph{ii)} can in principle be circumvented. Potter and Swendsen~\cite{Potter2013x} previously reported a violation of detailed balance on a broad set of update algorithms such as the sequential (typewriter) update and the checkerboard update. While in the context of nonequilibrium detailed balance is not a really meaningful concept, it led us to wonder whether the checkerboard dynamics that violates detailed balance might be different from the rsf dynamics satisfying detailed balance.
Reference~\cite{Potter2013x} further introduces variants of the algorithms which they prove do satisfy detailed balance. In the standard checkerboard algorithm the two sublattices are always updated in the same order, which they refer to as the \emph{01 cycle}. They claim that through a simple change, namely choosing at random which sublattice to update first, detailed balance can be reestablished. Note that still every spin is proposed to flip exactly once per sweep. This is referred to as \emph{mixed cycle}. For the double-checkerboard the order of the coarse checkerboards (red or blue first) is chosen at random once per sweep and the order of the fine checkerboard (lighter or darker sites first) is chosen at random and independently for each (red or blue) block and once per sweep. Reference~\cite{Potter2013x} proves that detailed balance is reestablished in this way in a general setting that entails many domain decomposition schemes including the double-checkerboard although not explicitly mentioned.

Lastly, at infinite temperature and using the Metropolis acceptance criterion $p_\text{acc}(\Delta E,T) = \min\{1,e^{-\Delta E / k_B T}\}$ all spin flip proposals are always accepted leading to problem \emph{iii)}. Consequentially, when every spin is proposed to flip once per sweep then one sweep consists of inverting the lattice. Hence, the system just switches between two states. Technically, this problem does only occur at infinite temperature but nonetheless causes large autocorrelation times at large finite temperatures. Reference~\cite{Weigel2021} comments on this issue but using the typewriter update instead of the checkerboard update. There spins are also proposed once per sweep (but in lexicographical order), and hence suffering the same fate as the checkerboard update at infinite temperature.
\subsection{$n$-fold way update}
At low temperatures, the problem of high rejection rates is well known (see, e.g., ch. 3.4.2 in Ref.~\cite{Puri_book}). Particularly in the case of $T=0$, many proposed spin flips will be rejected with 100\% certainty.

Already in the 1970s this problem was tackled by Bortz~\emph{et al.}~\cite{n-fold} with the introduction of the rejection-free $n$-fold way update algorithm. 
They recognized that instead of selecting each site with the same probability (as is the case in rsf) the chance of a site to be selected can be weighted by the acceptance probability of a spin flip which is then always accepted. This can be implemented very efficiently for systems with discrete spin variables and local interactions only by organizing spins in $n$ spin classes of equal acceptance probability.
In our case $n = 2 \times (6+1)$, as each site can take two values and the sum of its six nearest neighbor's spin values can take seven different values. 
To have a unit of time in MCS and not the number of successful flips one increments time after every flip by a geometrically distributed random variable with mean $N / \sum_i p_\beta([\Delta E]_i)$ where $p_\beta(\Delta E)$ being the usual acceptance probability if rsf were used~\footnote{One can easily convince oneself, that if the success probability of a single rsf spin proposal is $p_\text{succ}$, then the number of tries necessary to make a flip (irrespective of what site is flipped) follows a geometric distribution with mean $1/p_\text{succ}$}.

One way of improving the performance of $n$-fold way is to use continuous time steps instead of geometrically distributed ones, that is to increment time by the expected time. 
The difference between summing over many geometrically distributed variables and summing over their means is negligible on the time scales considered here which motivates this simplification.

The $n$-fold way update has been used to study coarsening before, see for example Refs.~\cite{n-fold, Soerensen1988, Blanchard2014}. In the case of the $T=0$ coarsening in $d=3$, Olejarz \emph{et al.}~\cite{olejarz2011zero} previously have used a very similar approach. Instead of complete rejection-free updates they pick at random from the set of allowed spin flips, i.e., they randomly propose to flip a spin from the set $\{\sigma_i | [\Delta E]_i \leq 0\}$. If $\Delta E < 0$ the move is immediately accepted and otherwise if $\Delta E = 0$, the move is accepted with 50\% probability.

\section{Supplementary Discussion: Numerical comparison of alternative spin updates}
\begin{figure}[t]
  \includegraphics{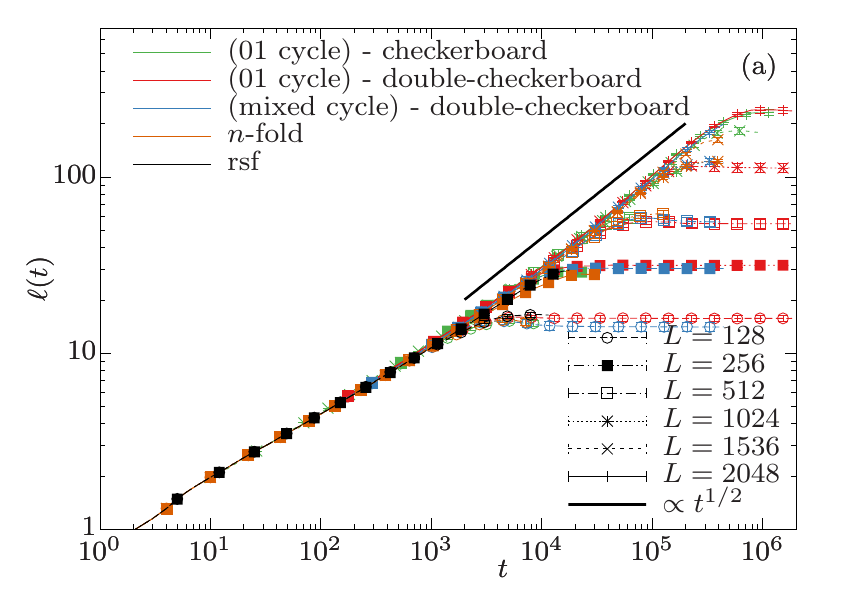}
  \includegraphics{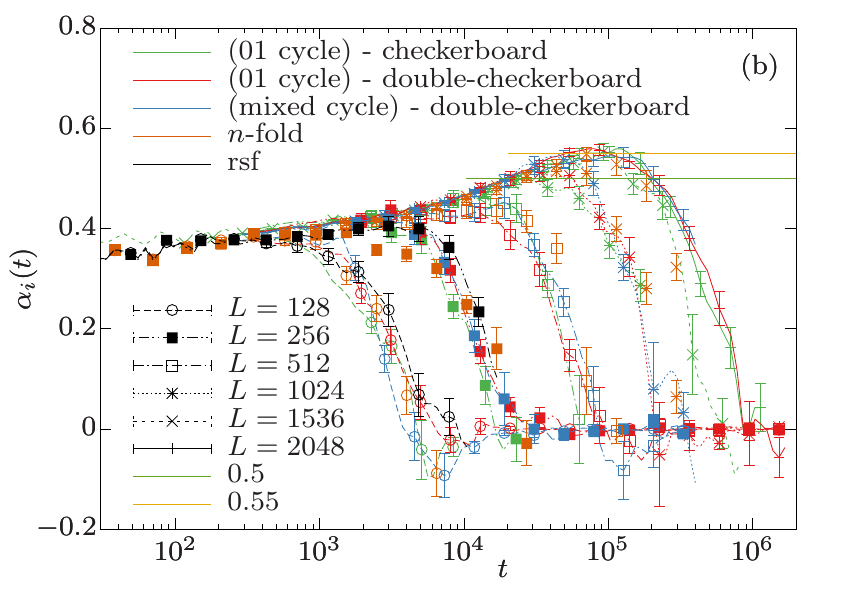}
  \caption{
    (a) Length scale $\ell(t)$ and (b) instantaneous growth exponent $\alpha_i(t)$ for various spin update methods obtained for quenches to $T=0$. In both plots the color of the line encodes the used method and the line type and symbol the system size. Note that for all methods (except rsf and $n$-fold) time is rescaled by a constant factor for better comparison (cf.\ Table~\ref{tab:factors}). For sake of readability, only every fifth data point is shown and lines are a guide to the eye connecting both shown and not shown points by straight lines. Error bars correspond to the standard error.
    }
    \label{figure:ell}
  \end{figure}

\begin{table}[b]
  \caption{Factors by which simulation times in Fig.~\ref{figure:ell} were multiplied to compare them. A higher number implies quicker dynamics.}
  \begin{tabular}{cl}
    \hline \hline
    \textbf{Method} & \textbf{Factor} \\ \hline
    (01 cycle) -- checkerboard & 1.75 \\
    (01 cycle) -- double-checkerboard & 1.75 \\
    (mixed cycle) -- double-checkerboard & 1.45 \\
    $n$-fold & 1 \\
    rsf & 1\\ \hline \hline
  \end{tabular}
  \label{tab:factors}
\end{table}

  In the following we present data obtained through the various different spin update schemes outlined above, where for the double-checkerboard we have tested two different cycles one of which preserves detailed balance. Every dataset is averaged over at least 40 independent realizations. The autocorrelation time is sensitive to the type of spin update employed. Similarly, the time scale in coarsening is affected by the chosen update. We used the dynamics from the rsf update as a base and then rescaled all other observation times by a factor specific to the used method resulting in a rescaled time scale $t$, see Table~\ref{tab:factors} for the factors. 

Figure~\ref{figure:ell}(a) shows the domain size and (b) the instantaneous growth exponent as a function of (rescaled) time~$t$. It can be seen that within error bars all the methods produce compatible results.

In particular all simulations show a growth exponent as large as $\simeq 0.55$ at late times. Using the $n$-fold way update, only linear system sizes of up to $1536$ were accessible to us, which also display the anomalous growth exponent larger than $0.5$. Hence, we can rule out that the larger exponent is a mere effect from the parallelized GPU update algorithms. Furthermore, this shows that the observation is in fact relevant in the thermodynamic limit and not just a finite-size effect. To check our $n$-fold way update implementation for correctness we used the rsf with which systems of linear size up to about $256$ are (easily) accessible. Note that due to the significant computing time required to simulate systems as large as $N=1536^3$ on CPUs we refrained from simulating past the onset of finite-size effects which is why the $L=1536$ line ends earlier than the other data sets displayed.
  
\begin{figure}[t]
  \includegraphics{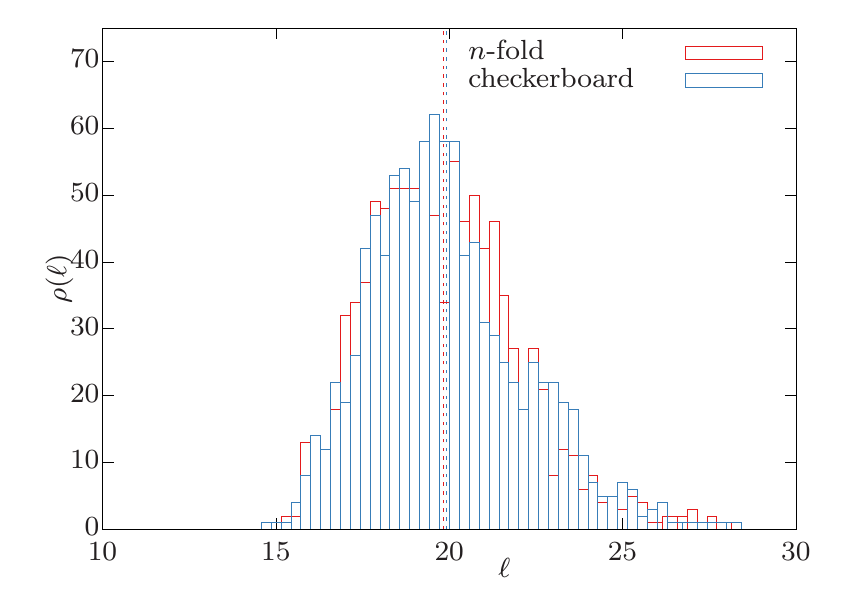}
  \caption{Empirically determined domain-size histogram for $L=256$ at rescaled time $t = 4000$ MCS using the $n$-fold way update and the (01-cycle) checkerboard decomposition. Data was obtained through 1000 independent runs for each method. Dashed vertical lines show the mean value of the two histograms, that is $19.8$ for $n$-fold way and $19.9$ for the checkerboard update.
  }
  \label{figure:ell_hist}
\end{figure}

  Finally, we run $1000$ simulations with system size $L=256$ once using the (01-cycle) checkerboard update and once using the $n$-fold way update. The first method is the one we used to obtain the data presented in the main text and the latter is known to obey the same dynamics as rsf Glauber dynamics. In Fig.~\ref{figure:ell_hist} we present a histogram of the measured domain sizes at a rescaled time $4000$ MCS (well before the onset of finite-size effects). The histograms appear to be in good agreement.

\section{Supplementary discussion: Comparison of different intercept values and effect on the calculated domain size}

In the following we repeat the measurement of the characteristic length scale $\ell(t)$ and vary the chosen intercept $c$ with $C(r,t)$. $\ell(t)$ is then the distance $r$ at which $C(r,t)$ intercepts $c$. We use the (01-cycle) checkerboard update (same as in the main text).

\begin{figure}[t]
  \includegraphics{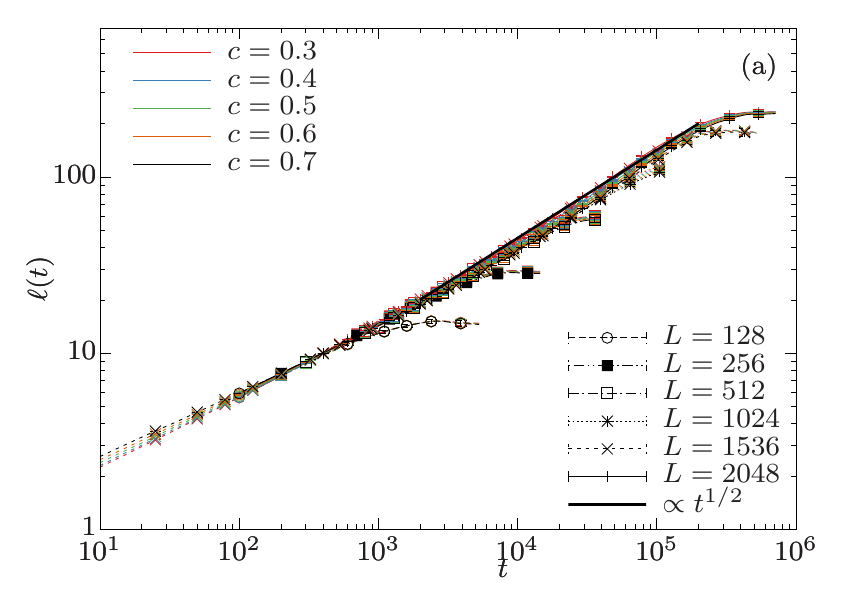}
  \includegraphics{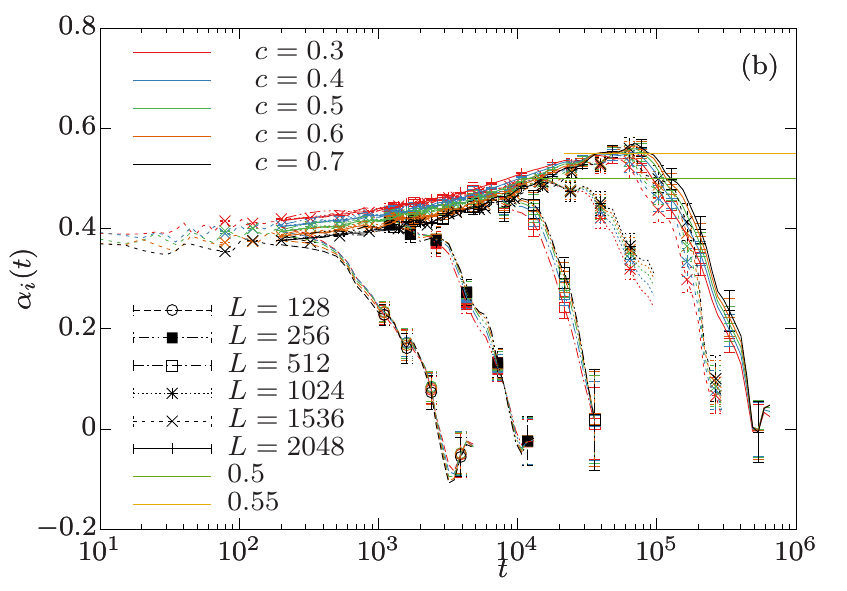}
  \caption{
    (a) Length scale $\ell(t)$ and (b) instantaneous growth exponent $\alpha_i(t)$ for various choices of the correlation-function intercept $c$. In both plots the color of the line encodes the used intercept and the line type and symbol the system size. Note that for all intercepts, the length scale is rescaled by a constant factor for better comparison (see Table~\ref{tab:interceptFactors}). For sake of readability, only every fifth data point is shown and lines are a guide to the eye connecting both shown and not shown points by straight lines. Error bars correspond to the standard error.
    }
    \label{figure:ellIntercept}
  \end{figure}

  \begin{table}[b]
    \caption{\textbf{Rescaling factors and saturation lengths for different intercepts.} The middle colum shows the factors by which the length scales in Fig.~\ref{figure:ellIntercept} were multiplied to compare them. The right column shows the length $\ell^{*}$ at which the not rescaled $\ell(t)$ estimate saturates.}
    \begin{tabular}{cll}
      \hline \hline
      \textbf{~Intercept $c$~} & \textbf{~Factor~} & \textbf{~Saturation Length $\ell^{*}$~} \\ \hline
      $0.3$ & ~0.61  & $~0.19\,L$\\
      $0.4$ & ~0.77  & $~0.15\,L$\\
      $0.5$ & ~1  & $~0.115\,L$\\
      $0.6$ & ~1.34  & $~0.086\,L$\\
      $0.7$ & ~1.90 & $~0.061\,L$\\ \hline \hline
    \end{tabular}
    \label{tab:interceptFactors}
  \end{table}

Figure~\ref{figure:ellIntercept} shows the different results for (a) $\ell(t)$ and (b) $\alpha_i(t)$. The characteristic length scales $\ell(t)$ for the various intercepts differ only by a constant factor and exhibit approximately the same $\alpha_i(t)$. To better show this, $\ell(t)$ is rescaled by this constant (independent of time and system size) for each intercept. Note, that this constant does not affect the obtained value for $\alpha_i$. The length scales obtained in this way are compatible with each other, thus confirming that the exact choice of $c$ is of little importance. Most importantly, note that all datasets for $L>1\,024$ and any intercept $c$ show $\alpha_i > 0.5$ at late times, thus confirming our observations in the main text. The times $t$ in Fig.~\ref{figure:ellIntercept} correspond to the not rescaled times. Hence, the data can readily be compared to Fig.~3 in the main text but times differ by a factor of $1.75$ to the ones in Fig.~\ref{figure:ell} in this supplement. The factors by which the estimates for $\ell$ from each intercept were multiplied such that the data sets collapse are compiled in Table~\ref{tab:interceptFactors}. Further, Table~\ref{tab:interceptFactors} shows the length scale $\ell^{*}$ at which $\ell(t)$ would saturate \emph{without rescaling}.

\section{Three-dimensional visualization}
\begin{figure*}[t]
  \begin{tabular}{cccc}
    \hline \hline
    \textbf{MC time} & \textbf{Cubic representation} & \textbf{Interface representation} & \textbf{Plane cut} \\ \hline
    \raisebox{1.9cm}{$t=9.5 \times 10^3$~\footnote{Instead of the full $1536^3$ snapshot we show a $1024^3$ subsection for the earliest time. This is on the one hand because otherwise the small features are hard to see and on the other hand because VMD appears to fail to handle the large number of ``atoms'' present in these early-time configurations. The cross-section shown in the third column has length $L=1536$ and the $1024$ square highlights the region visible in the 3D representations.}} & \includegraphics[width=0.3\textwidth]{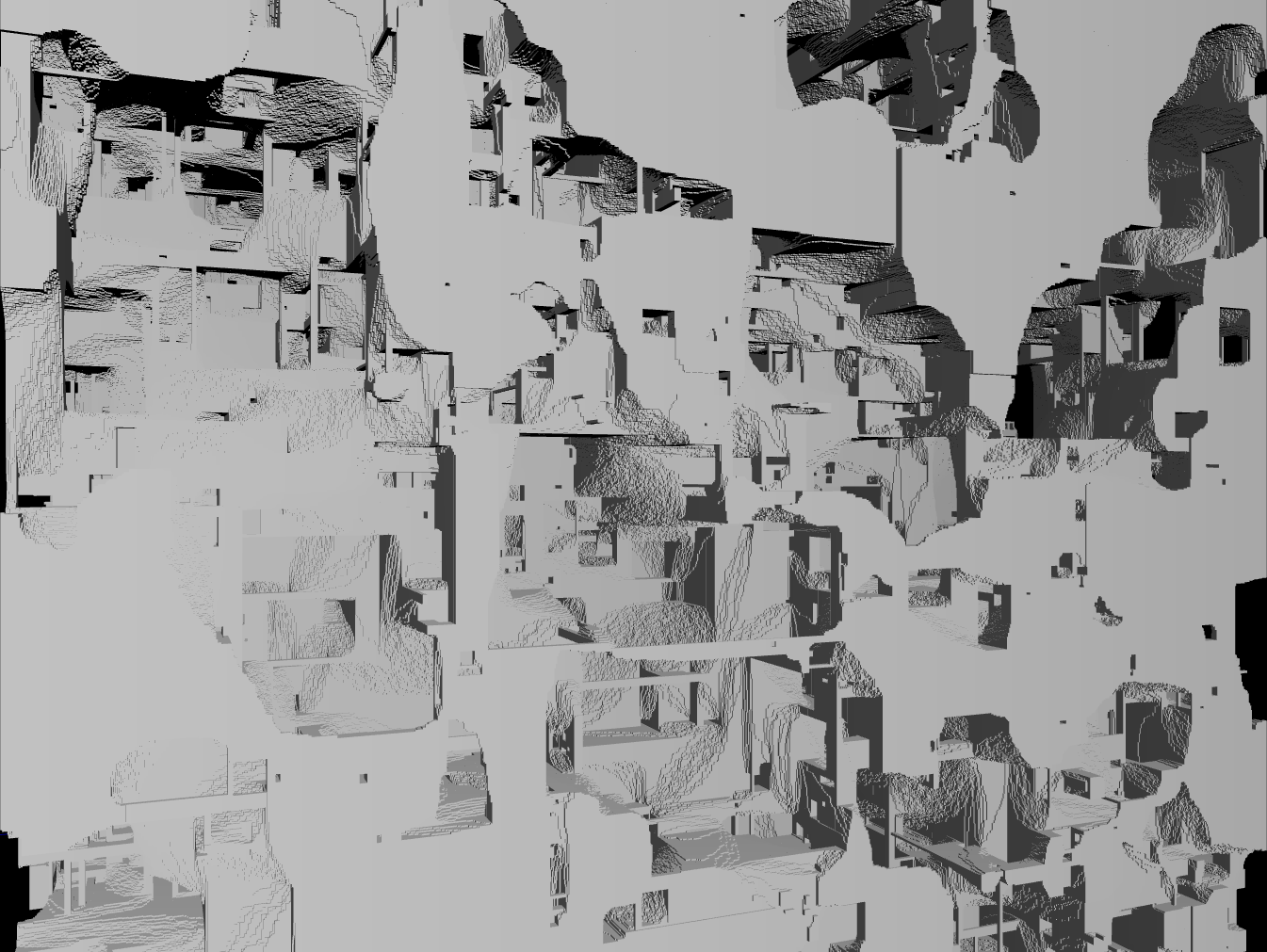} & \includegraphics[width=0.3\textwidth]{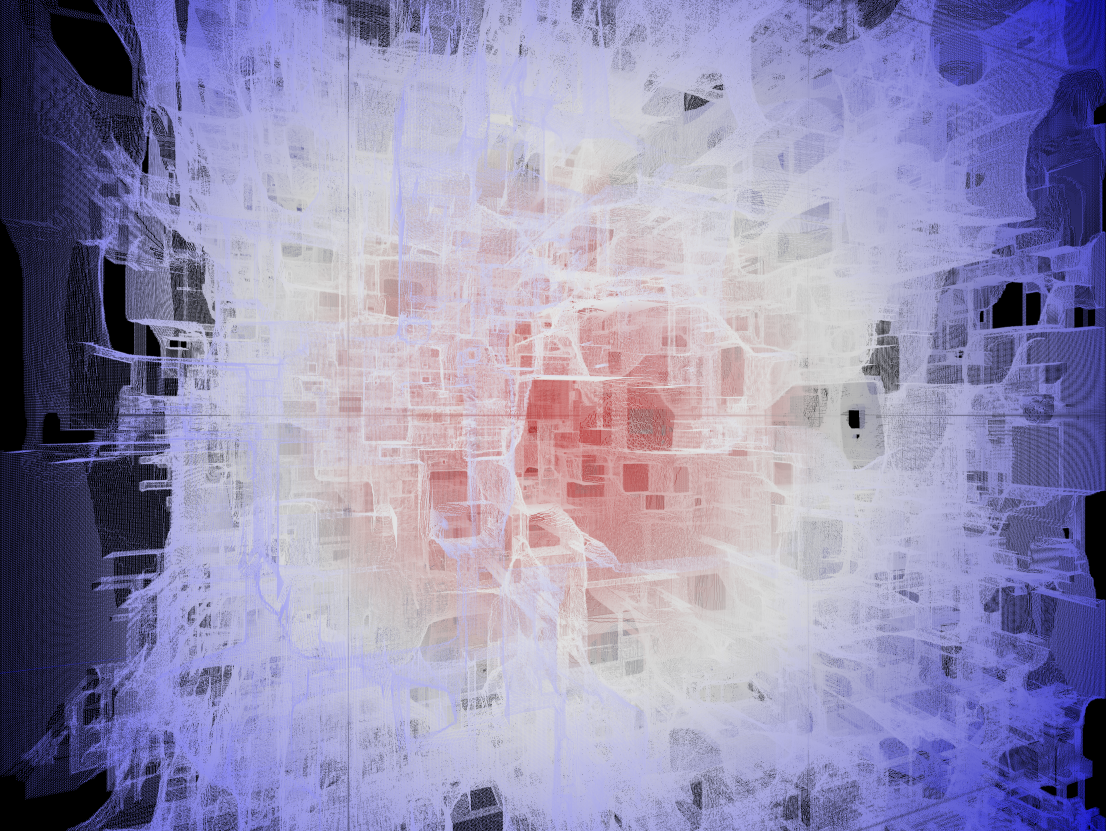} & \includegraphics[height=4.05cm]{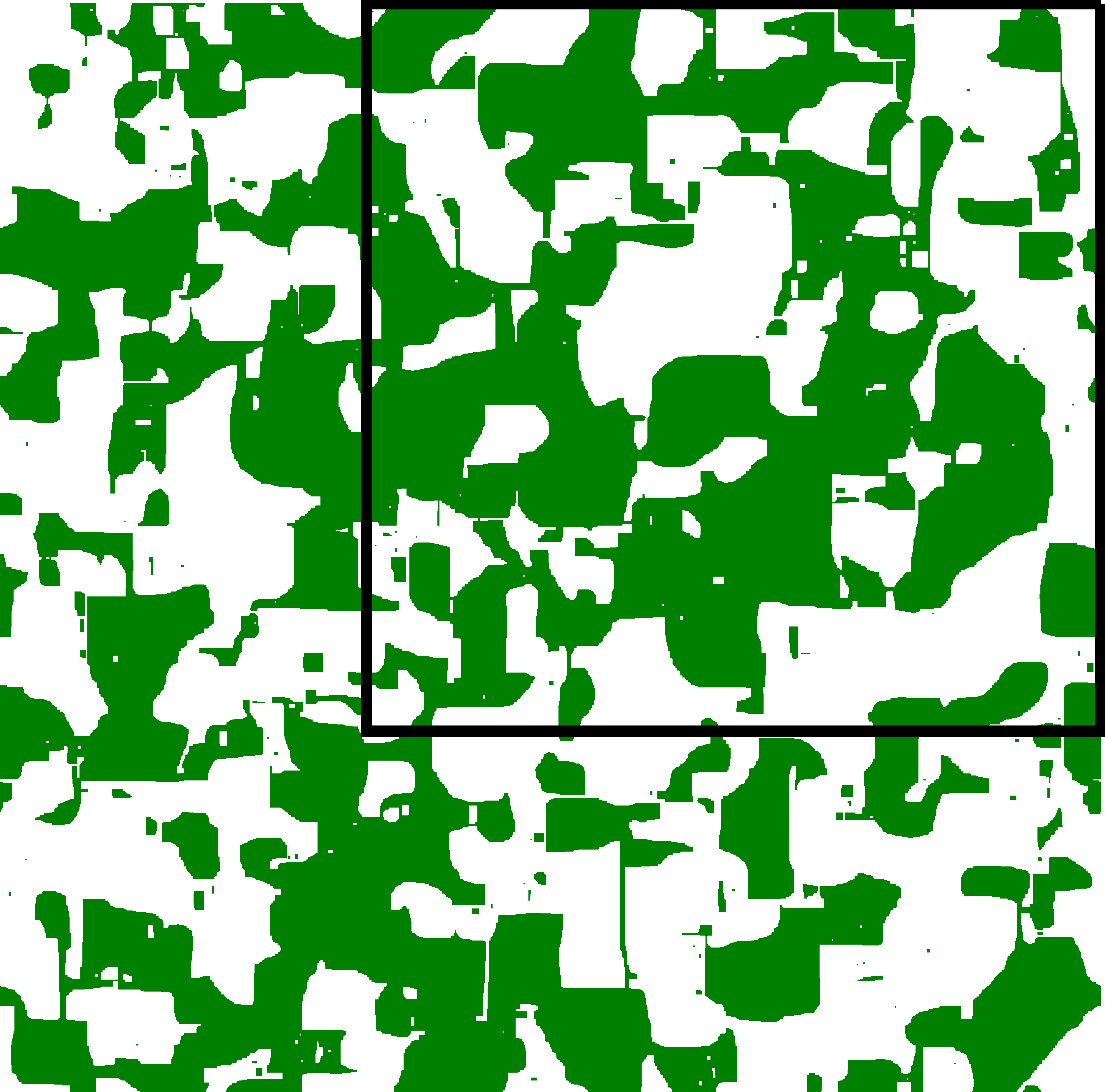} \\
    \raisebox{1.9cm}{$t=6.4\times 10^4$} & \includegraphics[width=0.3\textwidth]{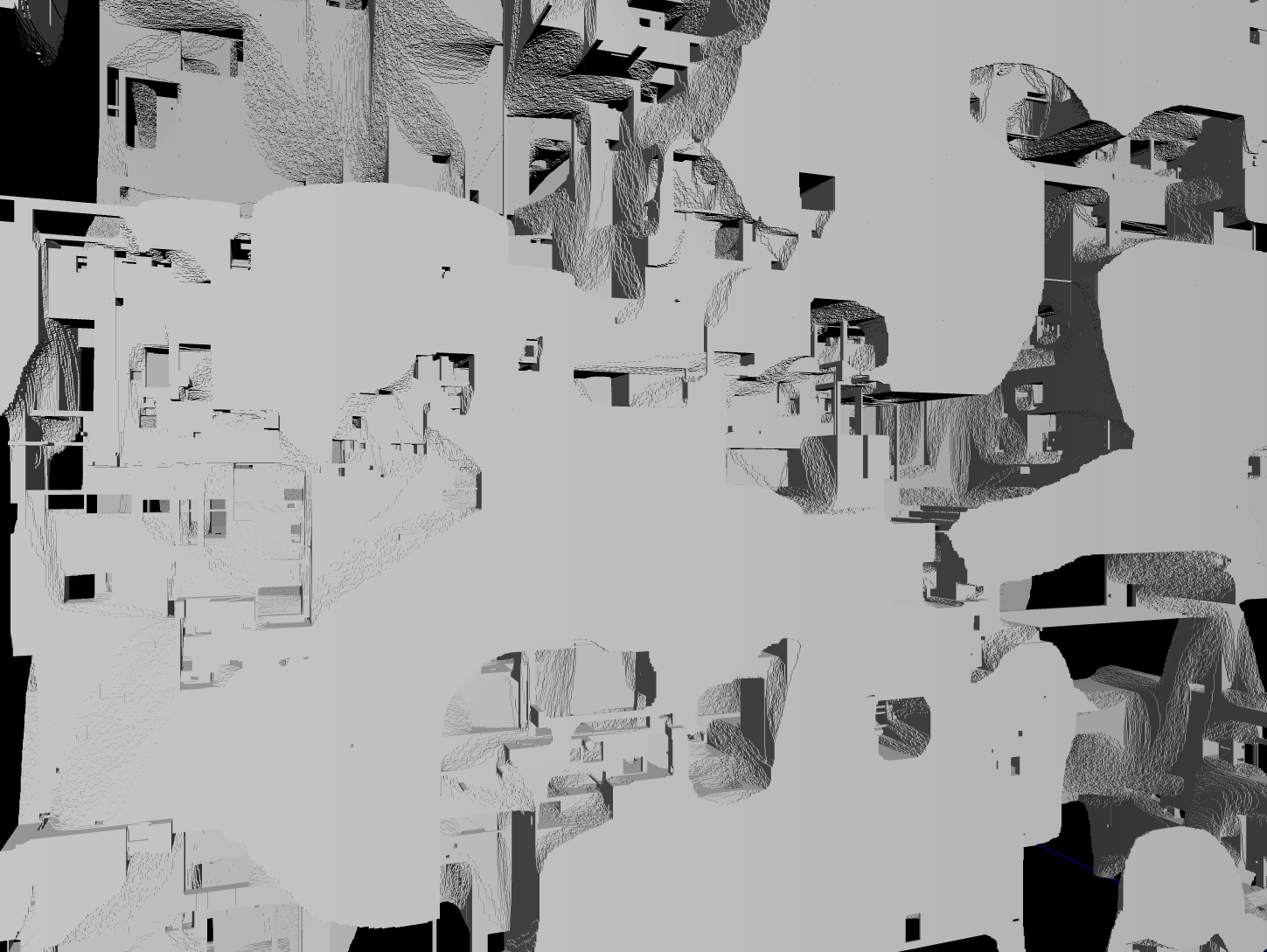} & \includegraphics[width=0.3\textwidth]{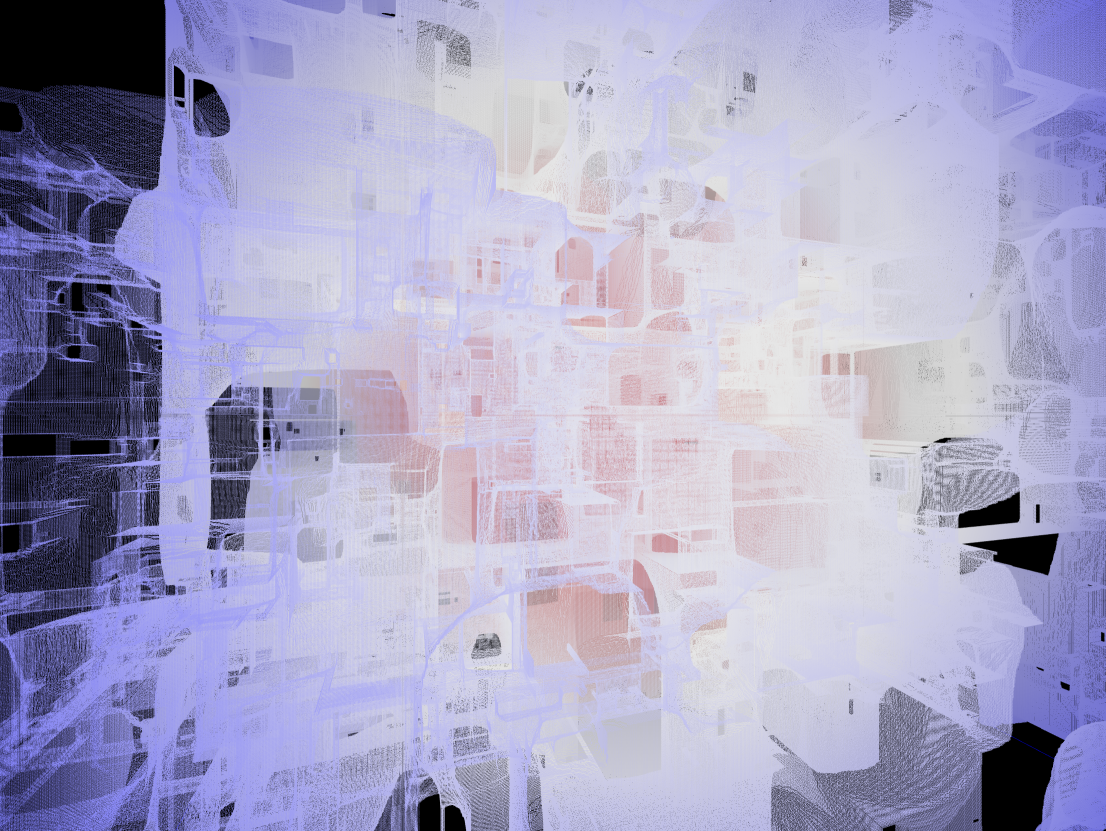} & \includegraphics[height=4.05cm]{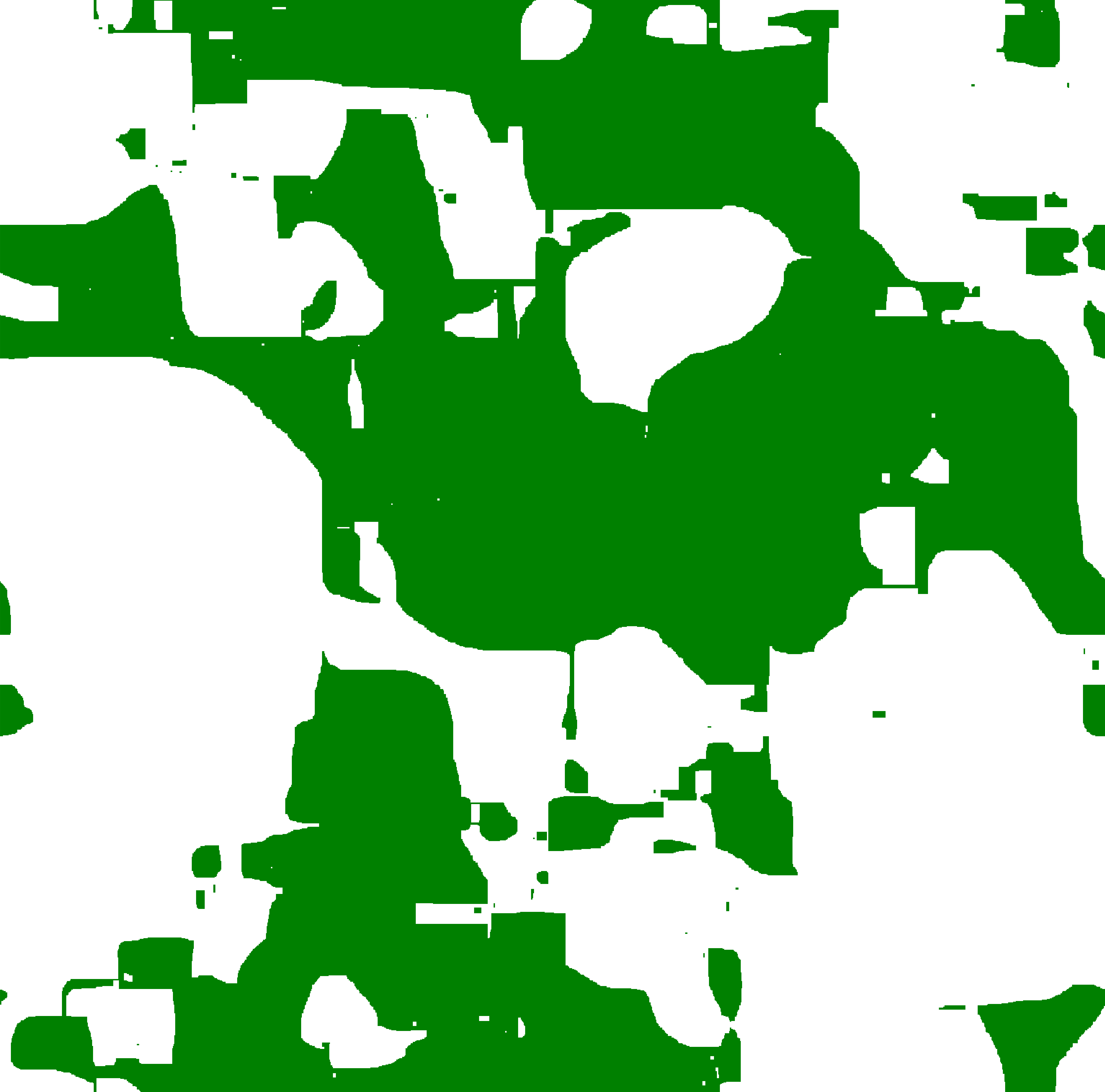} \\
    \raisebox{1.9cm}{$t=1.4\times 10^5$} & \includegraphics[width=0.3\textwidth]{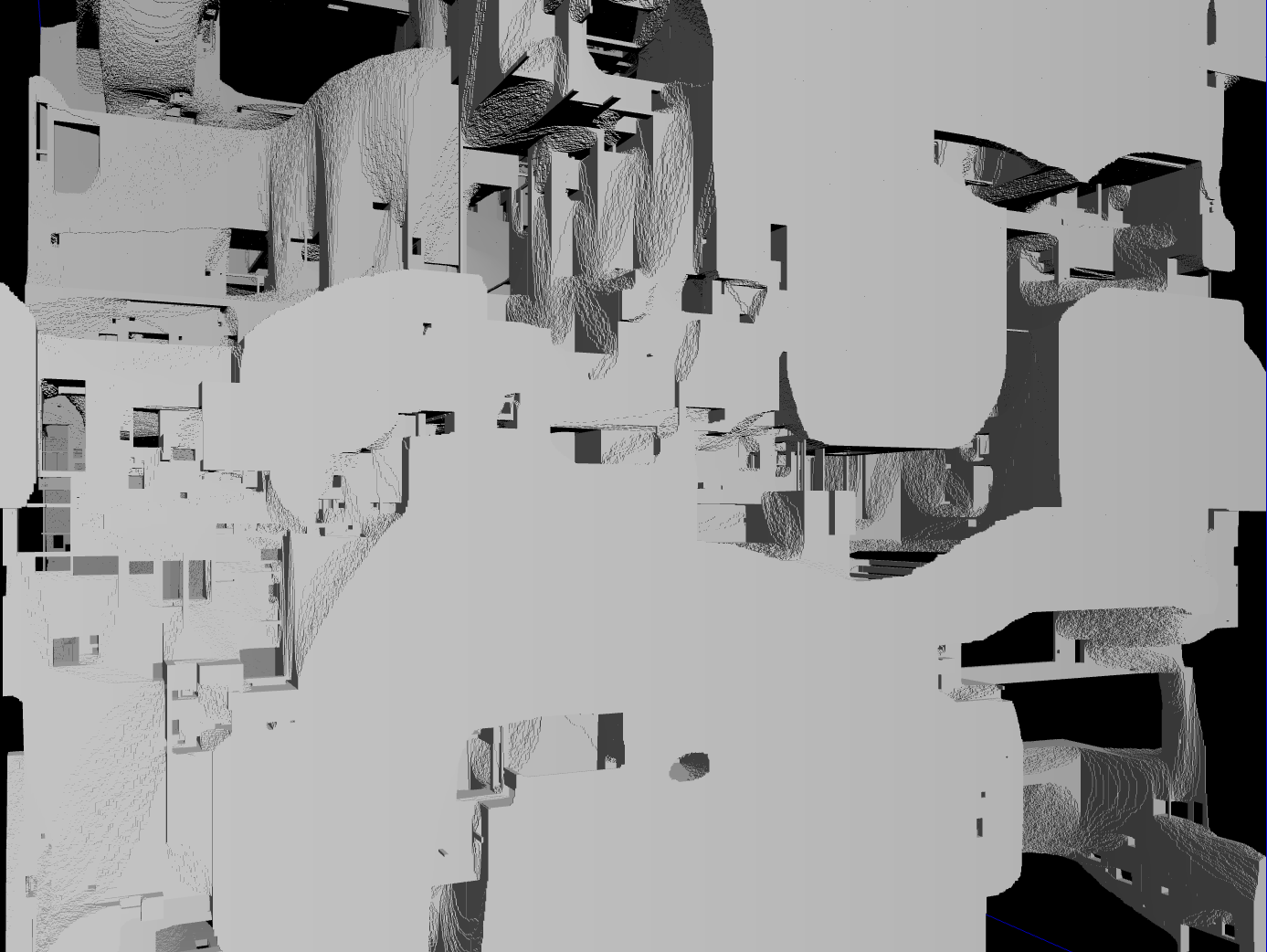} & \includegraphics[width=0.3\textwidth]{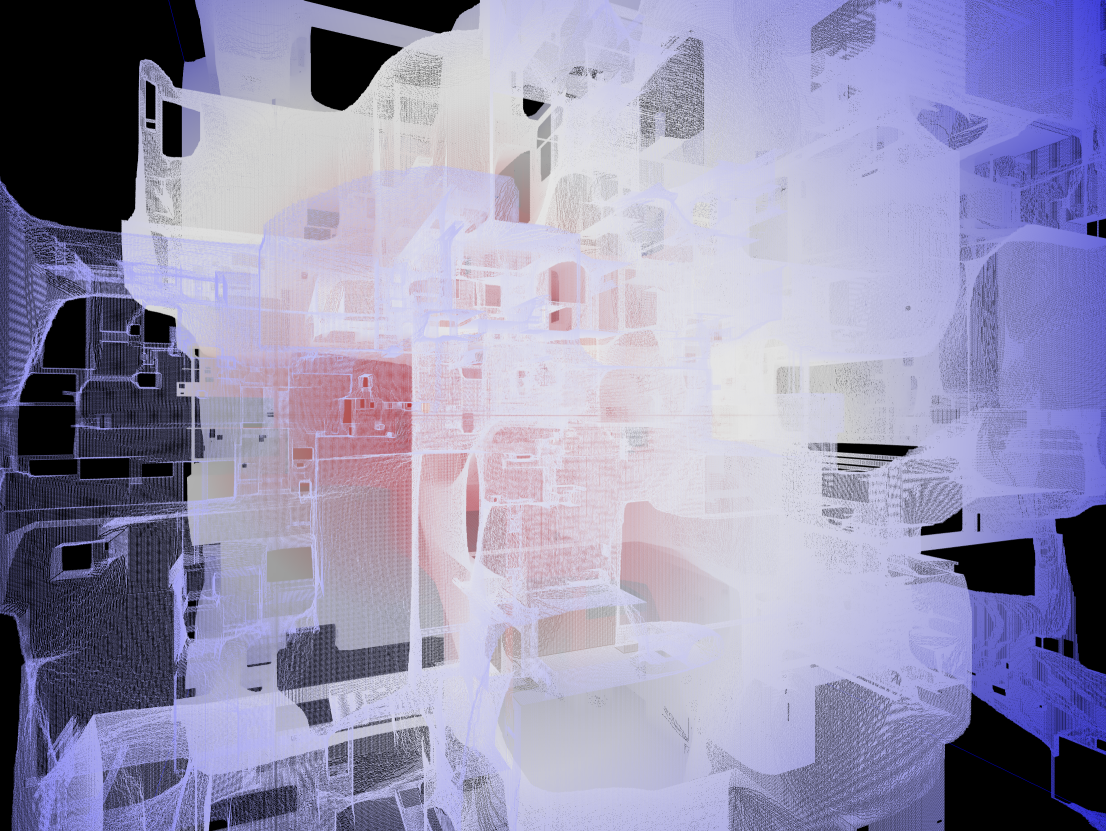} & \includegraphics[height=4.05cm]{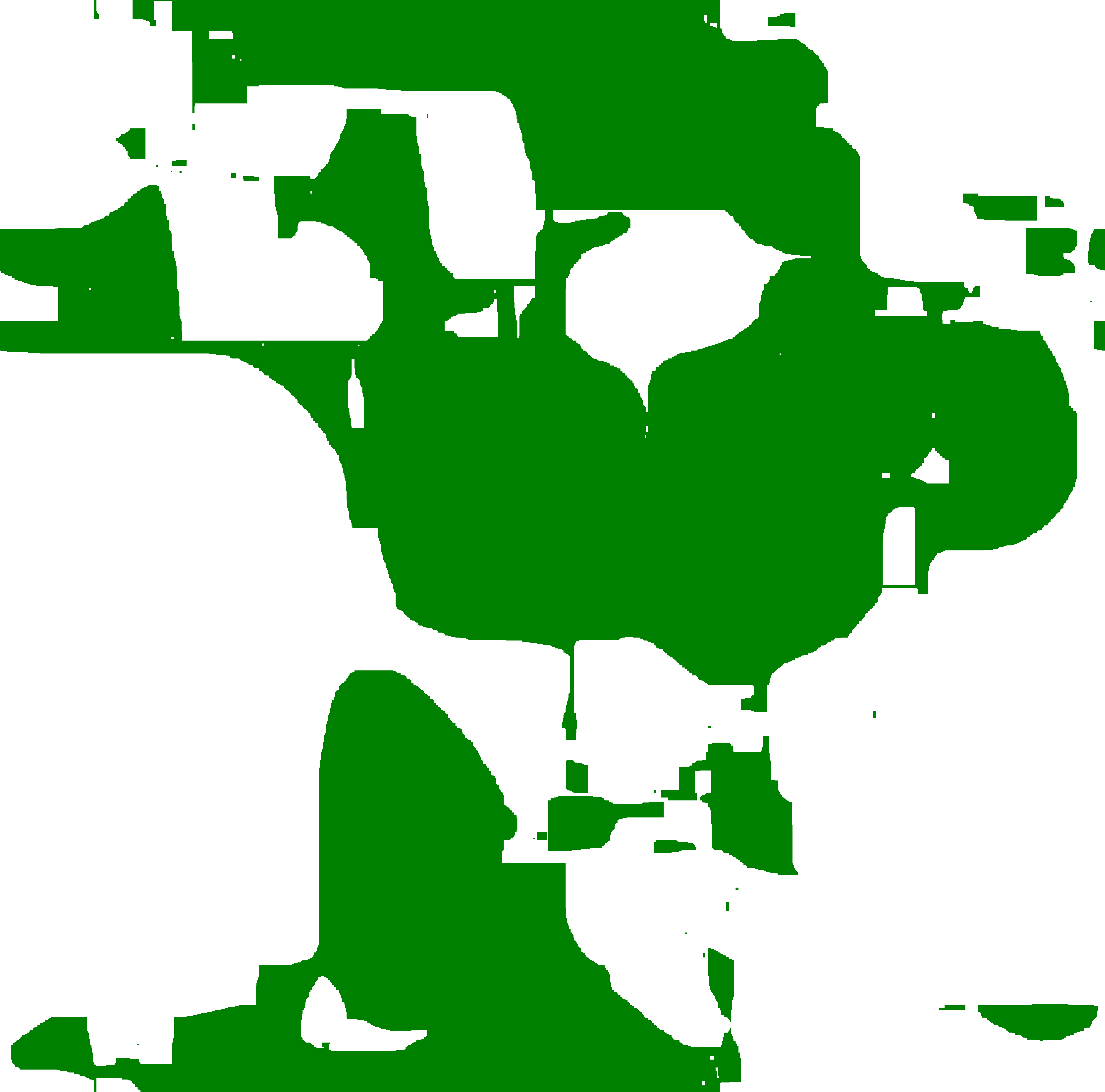} \\
    \raisebox{1.9cm}{$t=1.8\times 10^5$} & \includegraphics[width=0.3\textwidth]{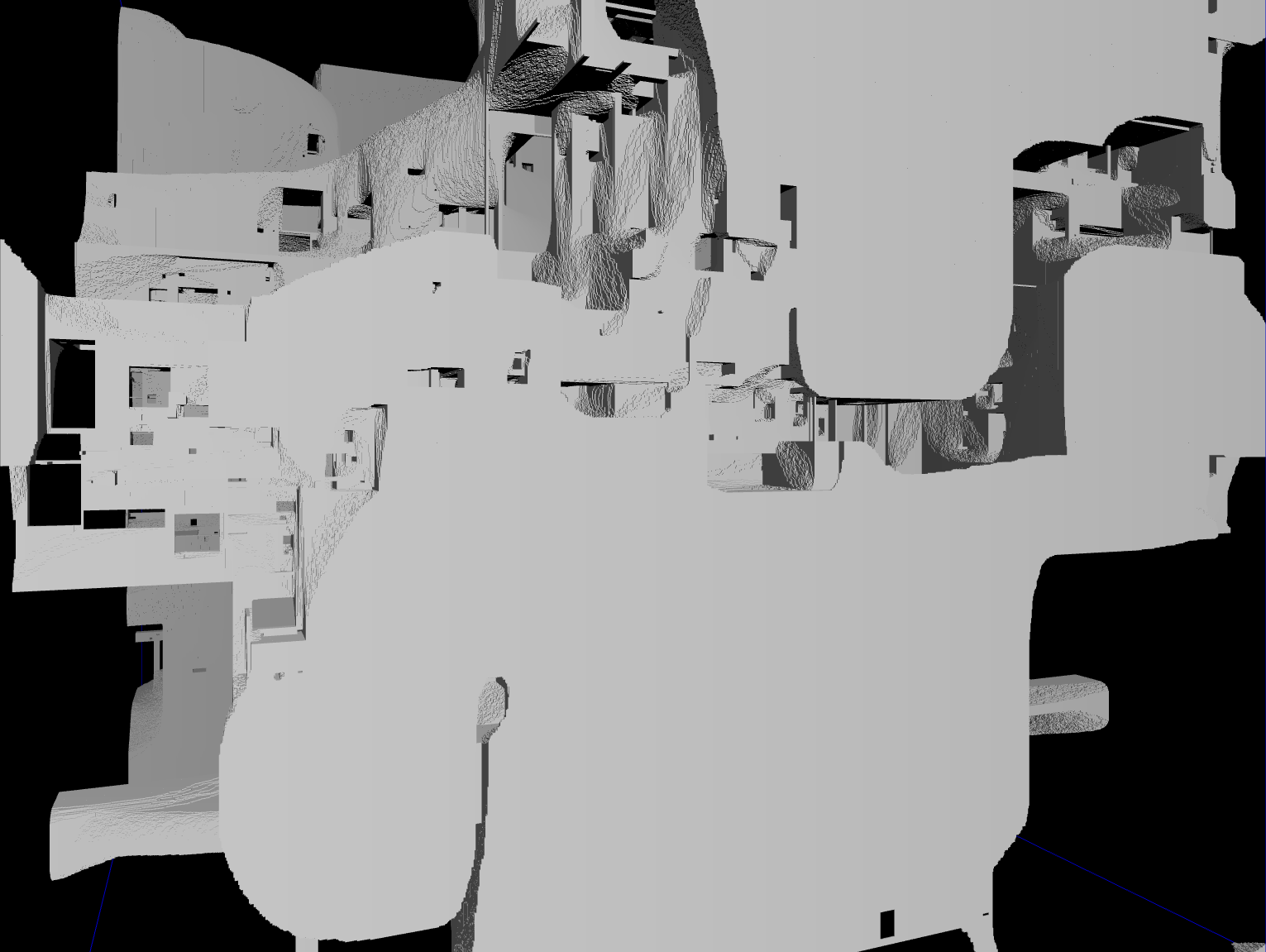} & \includegraphics[width=0.3\textwidth]{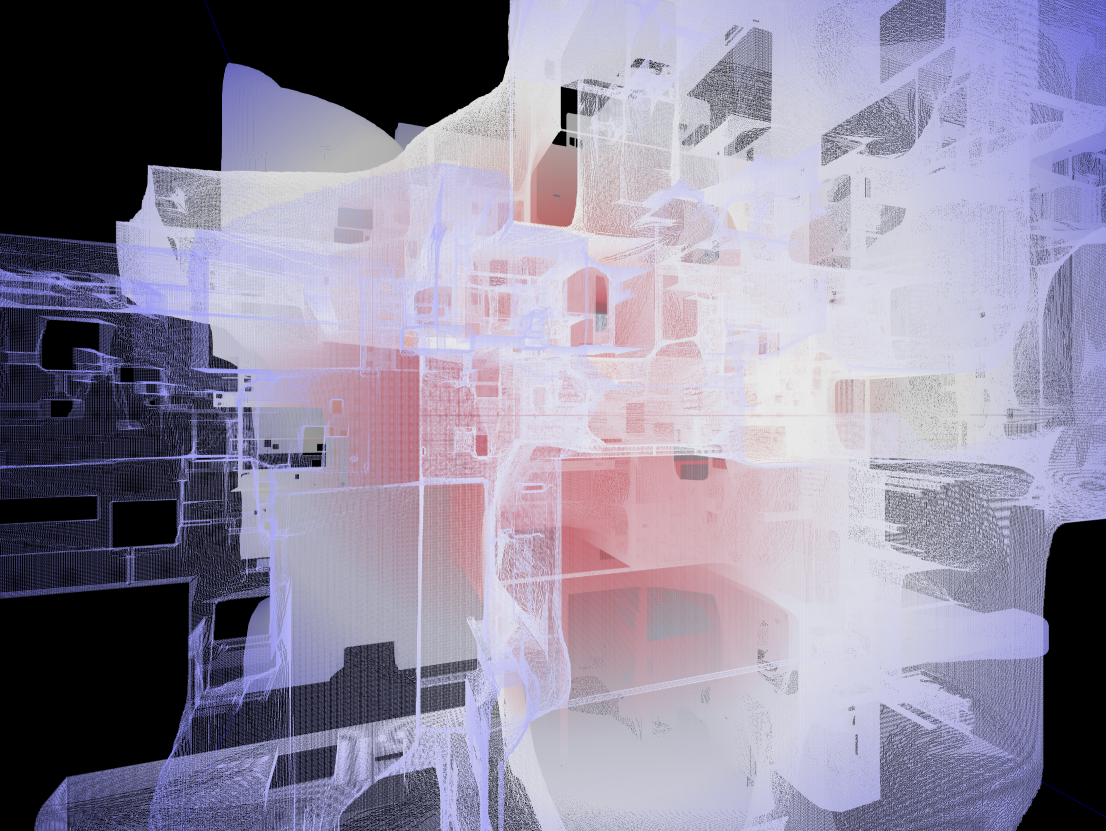} & \includegraphics[height=4.05cm]{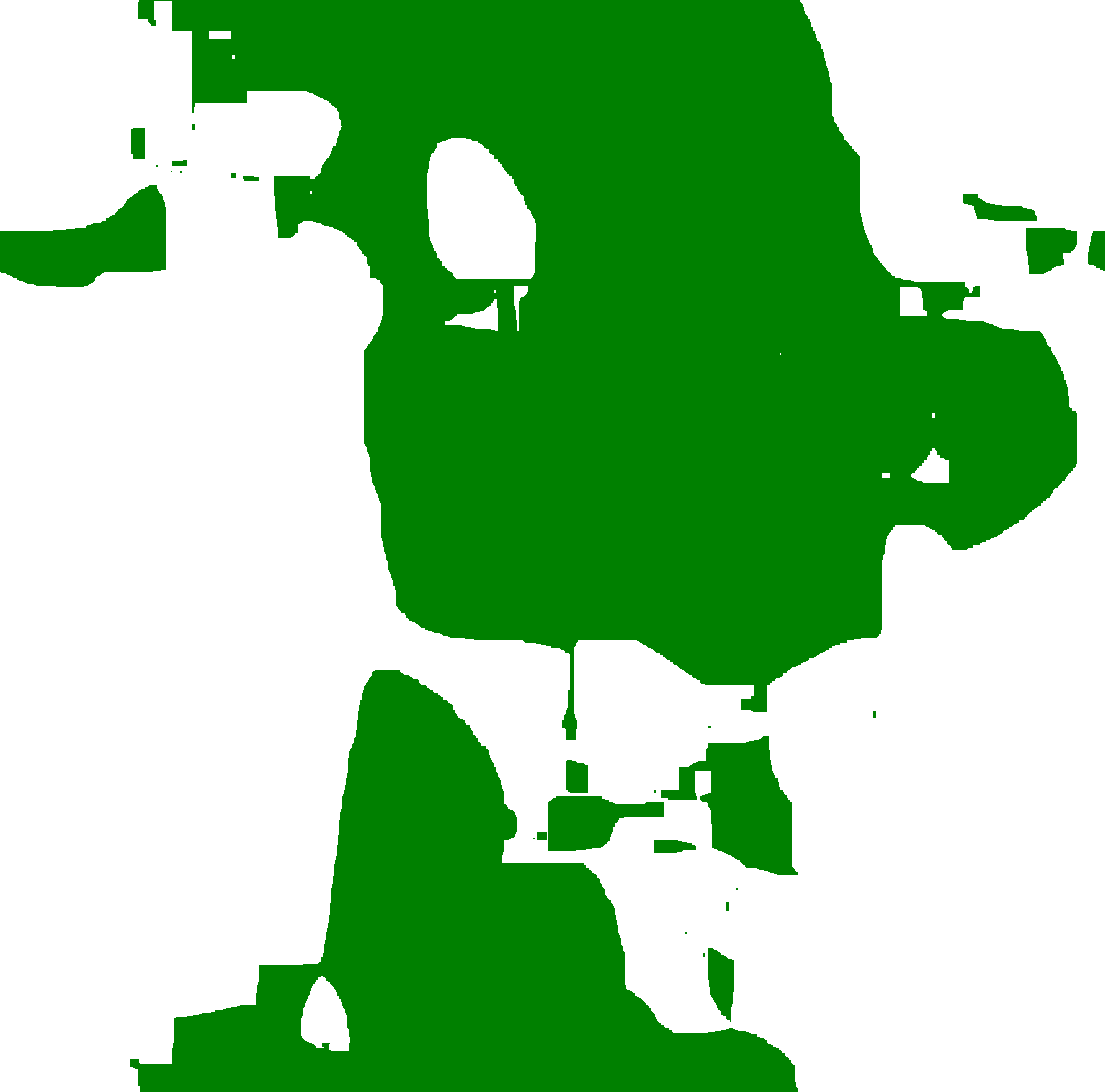} \\
    \raisebox{1.9cm}{$t=5.8\times 10^5$} & \includegraphics[width=0.3\textwidth]{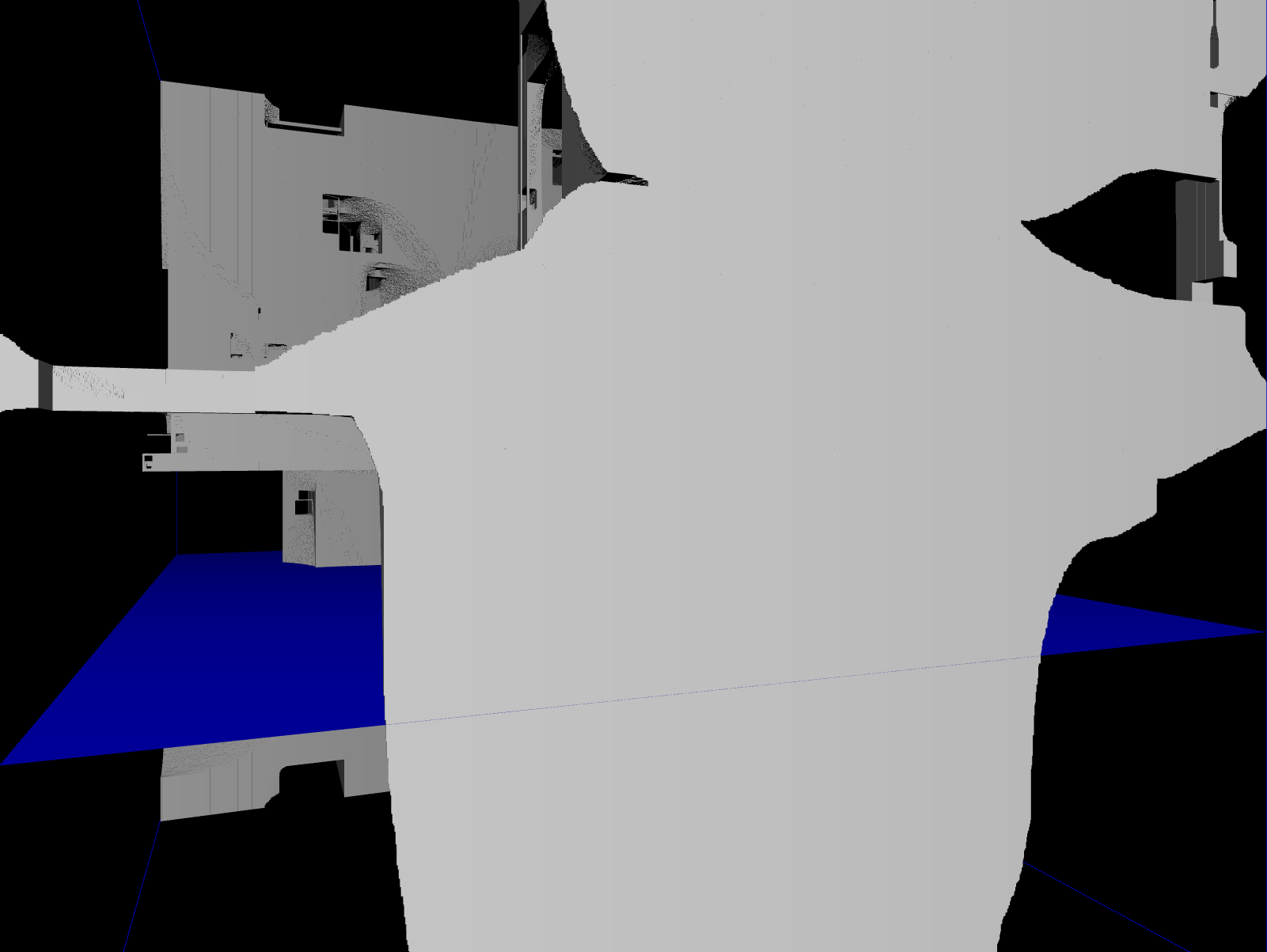} & \includegraphics[width=0.3\textwidth]{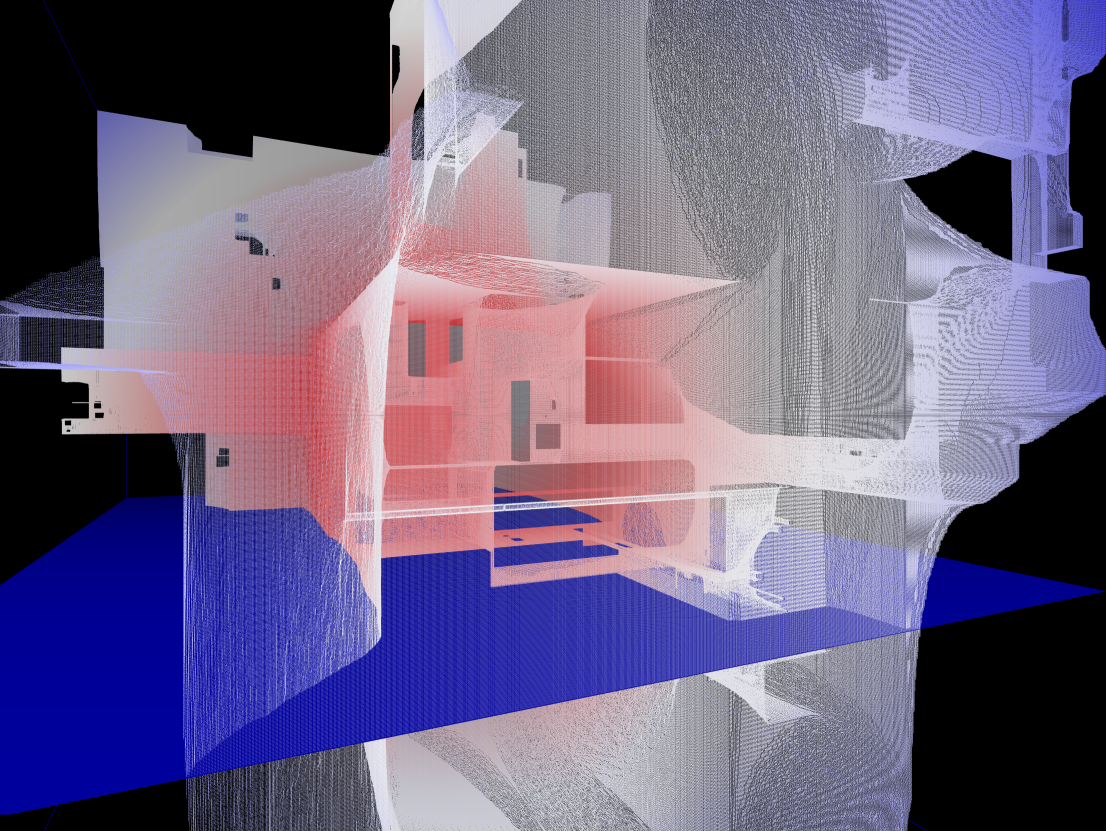} & \includegraphics[height=4.05cm]{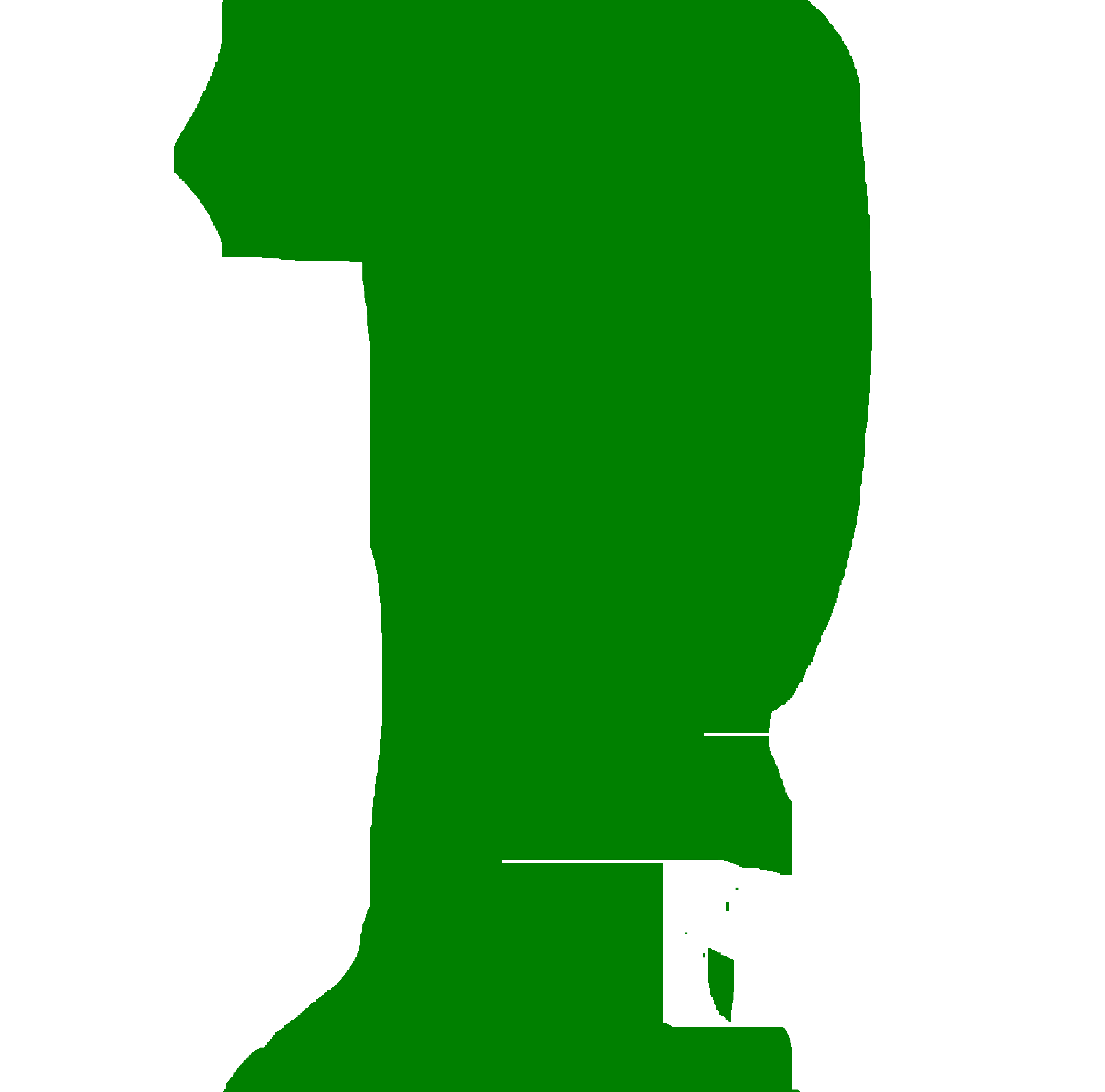} \\

     \hline \hline
  \end{tabular}
  \caption{Visualizations of snapshots obtained during a quench using $L=1536$. The plane cut is a horizontal slice from the three-dimensional structure, which for $t=5.8 \times 10^5$ is highlighted in the three-dimensional representations as the blue planes.}
  \label{tab:snapshots}
\end{figure*}

In Fig.~\ref{tab:snapshots} we show snapshots for different times in the evolution of the spin configuration from a single $n$-fold way simulation run with $L=1536$.
In this case we chose the $n$-fold way update instead of the checkerboard in an attempt to check whether any (obvious) structural differences can be seen. 
We make use of the software Visual Molecular Dynamics (VMD)~\cite{vmd} to illustrate the three-dimensional spin configurations.
In the cubic representation (left column) a spin is represented either by a cube or the lack thereof.
To make the plotting tractable, we choose to represent the minority direction as a cube (such that we always have to display less than $N/2$ cubes).
This representation directly shows the shape of typically encountered domains, but most of the inside of the lattice is hidden.
For a better illustration of the inside of the lattice, in the interface representation (center column) we only draw lines connecting the spins at the domain boundaries.
The color of the interfaces represents the distance from the center of the lattice (this is artificial, since periodic boundary conditions apply), where we employ a hot (red, center) to cold (blue, border) color palette.
This allows for a deeper look inside the lattice and highlights the nature of the domain boundaries, in particular the sponge-like structure.
Finally, we show two-dimensional cuts through the lattice (right column).

\newpage

%